\documentclass[fleqn,usenatbib]{mnras}

\usepackage{newtxtext,newtxmath}

\usepackage[T1]{fontenc}

\DeclareRobustCommand{\VAN}[3]{#2}
\let\VANthebibliography\thebibliography
\def\thebibliography{\DeclareRobustCommand{\VAN}[3]{##3}\VANthebibliography}

\usepackage{graphicx}	

\usepackage[mathcal]{euscript}
\usepackage{mathtools}
\usepackage{upgreek}
\usepackage{units} 
\newcommand{\e}{\ensuremath{\mathrm{e}}} 
\newcommand{\dif}{\,\mathrm{d}} 
\newcommand{\ode}[2]{\displaystyle\frac{\mathrm{d} #1}{\mathrm{d} #2}} 

\defcitealias{Benjamin:1997}{BD97}

\title[HVC survival and terminal velocity]{
\textbf{Gas accretion onto the Milky Way: high-velocity cloud survival and the revival of the terminal-velocity paradigm}
}

\author[M.~M.~Schulreich et al.]{
Michael M.~Schulreich,$^{1}$\thanks{E-mail: schulreich@astro.physik.tu-berlin.de}
Dieter Breitschwerdt$^{1}$
and J\"urgen Kerp$^{2}$
\\
$^{1}$Institut f\"ur Physik und Astronomie, Technische Universit\"at Berlin, Hardenbergstra{\ss}e 36, 10623 Berlin, Germany\\
$^{2}$Argelander-Institut f\"ur Astronomie, Auf dem H\"ugel 71, 53121 Bonn, Germany
}

\date{Accepted XXX. Received YYY; in original form ZZZ}

\pubyear{\the\year{}}

\begin{document}
\label{firstpage}
\pagerange{\pageref{firstpage}--\pageref{lastpage}}
\maketitle

\begin{abstract}
    The terminal-velocity paradigm has long been used to interpret the motion and infer the distances of high-velocity clouds (HVCs) accreting onto the Milky Way, yet its validity under realistic Galactic conditions remains uncertain.
    We investigate its dynamical limits by combining analytical modelling with three-dimensional hydrodynamical simulations of clouds moving through a stratified Milky Way halo. 
    We derive a generalized equation of motion including gravity, ram-pressure drag, a phenomenological mass-exchange model capturing mass loss and growth, and Bernoulli-driven cloud expansion. 
    Analytical solutions for constant-property clouds provide a reference framework, while the full evolution is assessed using simulations with adiabatic physics, radiative cooling, and thermal conduction.
    Terminal velocity is a local equilibrium but not a global attractor: dense clouds remain quasi-ballistic over most of their trajectories and approach terminal motion only shortly before reaching the Galactic disc. 
    Hydrodynamical effects further limit the paradigm. 
    In adiabatic flows, instabilities rapidly disrupt the cloud, rendering the terminal-velocity description inapplicable. 
    Radiative cooling instead promotes condensation and momentum loading, maintaining strong coupling to the background gas and restoring a terminal-velocity-like regime over extended periods, while thermal conduction mainly affects small-scale structure.
    Synthetic observables, including position--velocity diagrams, optical extinction, and soft X-ray emission, reproduce key features of observed HVCs such as velocity bridges and compression-driven emission. 
    We also provide direct measurements of the effective drag coefficient for infalling clouds, finding values of order unity but strongly time-dependent. 
    Overall, the terminal-velocity paradigm is a conditional description governed by cloud structure, mass exchange, and the ambient medium.
\end{abstract}




\section{Introduction}
 
The Milky Way, like other disc galaxies, did not form in a single monolithic collapse but has been assembled gradually, through the hierarchical merging of smaller systems and the continuous accretion of gas from its surroundings \citep[see e.g.][]{Mo:2010}.
Whereas merging dominated the early build-up, gas accretion is an ongoing process, as sustaining the Galaxy's star formation requires its gas reservoir to be continually replenished to offset the material locked up in stars.
Independent evidence for such present-day accretion comes from the metallicity distribution of long-lived stars in the solar neighbourhood, whose deficit of metal-poor G dwarfs relative to closed-box predictions -- the classical `G-dwarf problem' \citep{vandenBergh:1962,Schmidt:1963} -- is naturally explained by the infall of comparatively metal-poor gas \citep{Tinsley:1978}.
A significant fraction of this inflowing material is observed in the form of intermediate- and high-velocity clouds (IVCs and HVCs), neutral and partially ionized gas structures located in the Galactic halo and lower circumgalactic environment.
These clouds are defined by anomalous radial velocities with respect to the Local Standard of Rest (LSR), typically  $|\varv_{\mathrm{LSR}}|\gtrsim \unit[30]{km\,s^{-1}}$ for IVCs and $|\varv_{\mathrm{LSR}}|\gtrsim \unit[90]{km\,s^{-1}}$ for HVCs, with more refined classifications based on the deviation velocity relative to the maximum permitted Galactic disc velocity along a given line of sight (LOS) increasingly employed \citep[for comprehensive overviews, see e.g.][]{Wakker:1997,Wakker:2013,Richter:2006,Richter:2017,Kalberla:2009,Putman:2012}.

Since their discovery in \ion{H}{i} 21-cm emission \citep{Muller:1963}, IVCs and HVCs have been recognized as dynamically important components of the Galactic ecosystem.
Early studies already suggested a heterogeneous physical origin, including the infall of intergalactic gas \citep{Oort:1970}, the recycling of disc material through galactic fountain flows \citep{Shapiro:1976,Bregman:1980}, powered by the pressure of supernovae and cosmic rays \citep{Breitschwerdt:1991}, and tidal stripping during interactions with satellite galaxies, most prominently exemplified by the Magellanic Stream \citep{Mathewson:1974}.
As a result, IVCs and HVCs are observed to span a wide range of sizes and morphologies, from extended complexes and streams to compact, isolated clouds \cite[e.g.][]{Westmeier:2018}.
In modern theoretical frameworks, these formation pathways are understood as different manifestations of multiphase gas flows in the circumgalactic medium that regulate galactic accretion and feedback cycles \citep[see e.g.][and references therein]{Faucher-Giguere:2023}.
Cosmological simulations of Milky-Way-like galaxies now reproduce analogues of these clouds and their role in fuelling the disc down to the scale of individual structures \citep[e.g.][]{Lucchini:2024,Lucchini:2025}, underscoring the central importance of halo-cloud accretion for galaxy evolution. 
Resolving whether and how an individual cloud survives its passage through the halo, however, requires following the small-scale hydrodynamics that lies beyond the reach of such global models.

A long-standing difficulty in the study of IVCs and, in particular, HVCs is the determination of their distances.
Direct constraints from absorption-line spectroscopy towards halo stars provide reliable distance brackets for a subset of clouds, placing most IVCs at heights of $\sim$1--\unit[3]{kpc} and some HVC complexes at distances of several to tens of kiloparsecs from the Galactic plane.
However, such measurements require suitable background targets and are therefore observationally expensive and limited in sky coverage, leaving the distances of many HVCs poorly constrained \citep[see e.g.][]{vanWoerden:2004}.
This has motivated the development of \emph{dynamical} distance estimates, which aim to infer cloud distances from their observed velocities and column densities by modelling their motion through the Galactic halo.

A particularly influential dynamical framework was introduced by \citet[][hereafter \citetalias{Benjamin:1997}]{Benjamin:1997}, who modelled IVCs and HVCs as clouds moving through a stratified gaseous background under the combined action of gravity and ram-pressure drag.
In this picture, clouds approach a terminal velocity at which gravitational acceleration is balanced by ram-pressure drag, yielding a simple relation between velocity, column density, and height from the Galactic plane.
Such terminal-velocity arguments have been used both to estimate cloud distances and, inversely, to constrain the vertical structure of the halo \citep[e.g.][]{Putman:2012}.

Despite its conceptual clarity, the terminal-velocity picture relies on a number of simplifying assumptions, most importantly that clouds can be treated as coherent objects with fixed mass and geometry embedded in a prescribed environment.
Over the past two and a half decades, numerical simulations have demonstrated that clouds interacting with a hot halo are subject to Kelvin--Helmholtz (KH) and Rayleigh--Taylor (RT) instabilities, radiative cooling, and thermal conduction, which together drive mass loss, mass growth, and substantial morphological evolution \citep[e.g.][]{Santillan:1999,Gregori:2000,Quilis:2001,Vieser:2007,Heitsch:2009,Marinacci:2010,Shelton:2012,Scannapieco:2015,Brueggen:2016,Armillotta:2017,Gronke:2018,Gronke:2020,Li:2020,Lucchini:2024}.
A key insight to emerge from this body of work is that radiative cooling can offset the very instabilities that would otherwise destroy an infalling cloud. 
\citet{Marinacci:2010} showed that, above a critical ablation rate, cooling of gas stripped into the turbulent wake causes a comparable mass of ambient halo gas to condense back onto the cloud, so that KH ablation is balanced -- or outweighed -- by condensation and the cloud can survive and even grow as it moves through the hot halo; this condensation-driven transfer of halo gas is central to the galactic-fountain picture of disc replenishment \citep[for a review, see][]{Fraternali:2017}. 
Subsequent simulations established the same effect more quantitatively, with clouds surviving and accreting mass whenever the cooling time of the mixed gas is short compared with the time-scale on which instabilities disrupt them \citep[e.g.][]{Armillotta:2017,Gronke:2018,Gronke:2020}. 
Cloud survival is therefore not guaranteed but conditional, set by the competition between radiative cooling and instability-driven disruption.
These results raise a fundamental question: \emph{under which conditions does the concept of a terminal velocity remain a meaningful description of halo-cloud dynamics, and when does it break down due to cloud evolution and environmental coupling?}

In this paper, we address this question by systematically extending the \citetalias{Benjamin:1997} framework to account for time-dependent cloud properties, following earlier qualitative considerations regarding the importance of cloud survival for terminal-velocity applicability \citep{Benjamin:1999}, and incorporating recent insights into mass growth and momentum transfer \citep{Tan:2023}.
We develop a generalized analytical description of the cloud equation of motion (EOM) that includes physically motivated mass exchange driven by hydrodynamic instabilities, radiative cooling, and thermal conduction, together with realistic Galactic density and gravitational profiles and evolving cloud geometry.
In contrast to earlier treatments, clouds are modelled with spherical rather than cylindrical geometry, and the background medium is allowed to vary both with height above the Galactic plane and with Galactocentric radius.
Within this framework, terminal velocity does not generally act as a global attractor of the motion, but rather emerges as a local, time-dependent equilibrium that may or may not be approached depending on the interplay between initial conditions and cloud evolution.

For idealized cases in which the cloud properties are held fixed, the EOM admits analytical solutions expressible in quadrature for arbitrary initial velocities and vertically varying background profiles, thereby generalizing the terminal-velocity picture of \citetalias{Benjamin:1997}. 
To our knowledge, solutions of this form for quadratic drag in a gravitational field with spatially varying background conditions have not previously been derived and thus provide a useful reference framework for interpreting cloud trajectories. 
Once mass exchange, geometric evolution, or other time-dependent processes are included, the governing equations remain analytically defined but no longer admit such solutions and must instead be integrated numerically.

As part of this extension, we identify and quantify an additional dynamical effect not included in previous analytical terminal-velocity models: Bernoulli-driven deformation associated with the deflection of background gas around a moving cloud.
The resulting pressure gradients drive lateral expansion, increase the effective drag cross section, and introduce a dynamical feedback between cloud geometry and deceleration.
This coupling modifies the motion already during the early coherent phase and further constrains the applicability of simplified terminal-velocity arguments.
The underlying reason is that the interaction with the ambient medium introduces dissipation, such that the motion is no longer governed by conservative dynamics.

While the generalized analytical framework developed here captures the coupled effects of gravity, ram-pressure drag, mass exchange, and cloud deformation in a controlled manner, it remains necessarily limited once strong mixing, fragmentation, and thermal exchange dominate the cloud evolution.
Assessing the regime of validity of the analytical description therefore requires a fully numerical treatment.

To this end, we perform three-dimensional (3D) hydrodynamical simulations of clouds moving through a stratified Galactic background medium.
This `falling-cloud' setup differs from the widely used wind-tunnel (or cloud-in-wind) approach, in which a stationary cloud is exposed to a uniform inflow with fixed background properties \citep[e.g.][]{Scannapieco:2015,Armillotta:2017,Gronke:2018,Gronke:2020,Li:2020}.
While wind-tunnel experiments are well suited to isolating microphysics and achieving controlled parameter scans, they do not capture the simultaneous variation of gravity, stratification, and ambient pressure along a cloud trajectory.
In such setups, the relative velocity between the cloud and the ambient medium can adjust as the system evolves, such that momentum coupling between the phases may reduce the shear over time.
By contrast, for a cloud moving through a stratified background the relative velocity is continuously driven by gravity and does not relax in the same way, but instead tends to remain finite.
As a result, the cloud does not enter a regime of reduced shear, but remains persistently exposed to shear-driven instabilities; early examples combining wind-tunnel and free-fall-style setups already highlighted these differences \citep{Heitsch:2009}.
Our simulations include radiative cooling and, crucially, incorporate thermal conduction as an additional physical process that can compete with cooling and mixing in setting both the cloud lifetime and the persistence of a coherent cloud morphology.

The aim of this work is not to provide an exhaustive parameter study of halo-cloud properties, but to clarify the dynamical regimes in which terminal-velocity concepts remain applicable and to identify those in which they break down.
Our results have direct implications for dynamical distance estimates of HVCs, for the interpretation of decelerated clouds at the disc--halo interface, and for connecting observed cloud velocities to their evolutionary state.

Beyond their purely kinematic interpretation, the dynamical processes considered here are expected to leave observable imprints on the structure and emission properties of halo clouds.
Cloud deceleration and deformation, together with the stripping and mixing of cloud material in the wake, can produce intermediate-velocity gas that appears observationally as `velocity bridges' (VBs) in position--velocity space. 
Enhanced turbulent mixing may alter dust-to-gas ratios (DGRs) and optical extinction, and the interaction with the hot background medium can modify the soft X-ray (SXR) surface brightness of the surrounding gas.

The paper is organized as follows.
In Section~\ref{sec:dynamics} we develop a dynamical framework for the motion of IVCs and HVCs through the stratified Galactic background medium, beginning with a minimal EOM model and its analytical solution, and subsequently extending it to include mass exchange guided by the relevant physical time-scales. 
Cloud deformation is analysed separately using an analytical expansion model. 
The Galactic background medium and the numerical framework used to test these models are introduced at the end of that section.
Section~\ref{sec:results} presents the analytical results for idealized cases together with the outcomes of the hydrodynamical simulations.
In Section~\ref{sec:discussion} we discuss the implications of our findings for IVC and HVC dynamics, addressing both observational connections and the limitations of the present study.
Our conclusions are summarized in Section~\ref{sec:conclusions}.

\section{Dynamics of IVCs and HVCs}
\label{sec:dynamics}

\subsection{EOM and baseline assumptions}
\label{sec:eom}

We model an IVC or HVC as a coherent structure undergoing bulk vertical motion through the Galactic background medium under the combined influence of gravity, ram-pressure drag, and mass exchange.
The EOM for such a cloud can be written as
\begin{equation}
    \begin{split}
        &M_\mathrm{cl}(t) \ode{\varv_\mathrm{cl}}{t} + \varv_\mathrm{cl}(t) \ode{M_\mathrm{cl}}{t} \\
        &\quad = \pm \frac{1}{2} C_\mathrm{d}(t) A_\mathrm{cl}(t) \rho_\mathrm{bg}(z(t);R) \left[\varv_\mathrm{cl}(t)-\varv_\mathrm{bg}(z(t);R)\right]^2 \\
        &\qquad - \left[M_\mathrm{cl}(t) - \rho_\mathrm{bg}(z(t);R) V_\mathrm{cl}(t)\right] \varg(z(t);R),
    \end{split}
    \label{eq:eom_masschange}
\end{equation}
where $M_\mathrm{cl}$, $V_\mathrm{cl}$, $A_\mathrm{cl}$, $\varv_\mathrm{cl}\equiv \mathrm{d}z/\mathrm{d}t$, and $C_\mathrm{d}$ denote the cloud's mass, volume, projected frontal area, vertical velocity, and (ram-pressure) drag coefficient, respectively.
These quantities may in general vary over time $t$ due to mass evolution, deformation, or changing flow conditions.
The background medium is characterized by its density $\rho_\mathrm{bg}$, bulk vertical velocity $\varv_\mathrm{bg}$, and vertical gravitational acceleration $\varg$, which vary with Galactic height $z$ (and thus implicitly with $t$).

Throughout this work, the cloud motion is assumed to be purely vertical at a fixed Galactocentric radius $R$, so that all background quantities are evaluated along the cloud trajectory and depend only on $z$; the parameter $R$ is therefore treated as implicit unless explicitly stated otherwise (for a discussion, see Section~\ref{sec:limitations}).
Bulk background flows are neglected ($\varv_\mathrm{bg}=0$), and for brevity the subscript `cl' is omitted from the velocity unless ambiguity may arise. 

It is convenient to introduce three auxiliary coefficients: the deceleration parameter
\begin{equation}
    \alpha(t) \equiv \frac{C_\mathrm{d}(t) A_\mathrm{cl}(t) \rho_\mathrm{bg}(z(t))}{2 M_\mathrm{cl}(t)},
    \label{eq:decpar}
\end{equation}
the specific mass-exchange rate
\begin{equation}
    \beta(t) \equiv \frac{1}{M_\mathrm{cl}(t)} \ode{M_\mathrm{cl}}{t},
    \label{eq:specmassexrate}
\end{equation}
and the cloud-to-background density contrast
\begin{equation}
    \chi(t) \equiv \frac{\rho_\mathrm{cl}(t)}{\rho_\mathrm{bg}(z(t))},
    \label{compratio}
\end{equation}
where $\rho_\mathrm{cl}=M_\mathrm{cl}/V_\mathrm{cl}$ denotes the volume-averaged cloud density.\footnote{If the cloud is in pressure equilibrium with its surroundings, $p_\mathrm{cl}=p_\mathrm{bg}$, the density contrast can equivalently be written as a temperature contrast, $\chi=T_\mathrm{bg}/T_\mathrm{cl}$.} 
With these definitions, equation~\eqref{eq:eom_masschange} takes the compact form
\begin{equation}
    \ode{\varv}{t} = \pm \alpha(t) \varv^2(t) - \beta(t) \varv(t) - \left[1 - \chi^{-1}(t)\right] \varg(z(t)).
    \label{eq:eom_masschange_compact}
\end{equation}
The upper sign applies to motion towards decreasing $z$ ($\varv<0$), the lower sign to motion towards increasing $z$ ($\varv>0$); this convention is used consistently throughout the paper wherever $\pm/\mp$ notation appears.\footnote{The term $\pm \alpha \varv^2$ can equivalently be written as $-\alpha \varv |\varv|$; the present form is adopted for analytical convenience.}

The first term on the right-hand side of equation~\eqref{eq:eom_masschange_compact} represents ram-pressure drag, which always acts opposite to the direction of motion.
The second term describes inertial effects associated with mass exchange: mass loss ($\beta<0$) enhances motion in the current direction, whereas mass gain ($\beta>0$) leads to deceleration.
The final term corresponds to the effective vertical acceleration due to gravity reduced by buoyancy and directed towards the Galactic mid-plane, with $\varg>0$ for $z>0$ and $\varg<0$ for $z<0$.
In the regime relevant for most of this study, the cloud remains much denser than its surroundings ($\chi\gg1$), so that buoyancy provides only a minor correction and the effective acceleration approaches $-\varg$.

Although $V_\mathrm{cl}$, $A_\mathrm{cl}$, and $C_\mathrm{d}$ may evolve due to ablation, deformation, or changing flow conditions, we adopt constant values for these quantities in the baseline model in order to isolate the effects of gravity, stratification, and mass exchange. 
This assumption is partially relaxed in Section~\ref{sec:bernoulli}.
The drag coefficient depends on cloud geometry and flow regime and typically spans a broad range, from $C_{\rm d}\sim0.1$ for streamlined configurations to values of order unity or larger for blunt or irregular bodies \citep{Hoerner:1965}.

To connect the baseline model to observable cloud properties, the initial mass must be specified in terms of measurable quantities.
For a homogeneous, pure-hydrogen cloud with either spherical ($j=1$) or cylindrical ($j=0$) geometry, the initial mass can be written as
\begin{equation}
    M_{\mathrm{cl},0} = (2/3)^j \bar{m} A_\mathrm{cl} f_\mathrm{cl}^{-1} N_\ion{H}{i},
    \label{eq:initcloudmass}
\end{equation}
where $A_\mathrm{cl}=\uppi R_\mathrm{cl}^2$, with $R_\mathrm{cl}$ the characteristic cloud radius.
$N_\ion{H}{i}$ and $N_\ion{H}{ii}$ denote the peak column densities of neutral and ionized hydrogen, respectively, from which the neutral fraction is defined as $f_\mathrm{cl}\equiv N_\ion{H}{i}/\left(N_\ion{H}{i}+N_\ion{H}{ii}\right)$.
The mean mass per particle is $\bar{m} = m_\mathrm{p} / \left(2 - f_\mathrm{cl}\right)$, with $m_\mathrm{p}$ the proton rest mass.

\subsection{Analytical solution and kinematic regimes}
\label{sec:analytics}

\subsubsection{Terminal velocity}

At any given height, an instantaneous terminal velocity can be defined by the condition of vanishing acceleration ($\mathrm{d}\varv/\mathrm{d}t=0$), corresponding to local force balance between gravity, ram-pressure drag, and inertial effects associated with mass exchange.
Applying this condition to equation~\eqref{eq:eom_masschange_compact} yields the local stationary solutions
\begin{equation}
    \varv_\mathrm{T} = \frac{\pm\beta \mp \sqrt{\beta^{2} \pm 4\alpha \left(1-\chi^{-1}\right) \varg}}{2\alpha}.
    \label{eq:vterm_general}
\end{equation}
A real-valued terminal velocity requires that the corresponding discriminant be non-negative. 
In addition, the physically admissible solution must have the correct sign for the assumed direction of motion.

Because $\alpha>0$ and $\chi>1$ for overdense clouds, this expression immediately implies a simple physical result: a terminal velocity directed towards the Galactic mid-plane always exists, independent of the sign of $\beta$. 
In contrast, a stationary solution directed away from the mid-plane exists only if the discriminant is non-negative and $\beta<0$, i.e.~if sufficiently strong mass loss offsets the combined effects of gravity and ram-pressure drag.

In the dense-cloud limit ($\chi\gg1$) and in the absence of mass exchange ($\beta=0$), equation~\eqref{eq:vterm_general} admits real-valued terminal solutions only for motion towards the Galactic mid-plane and reduces to
\begin{equation}
    \varv_\mathrm{T} = -\mathrm{sgn}(z) \sqrt{\frac{|\varg|}{\alpha}}.
    \label{eq:vterm_simple}
\end{equation}
As both $\alpha$ and $\varg$ generally vary with height in a stratified Galactic background medium, the terminal velocity evolves along the trajectory rather than remaining constant.

\subsubsection{Analytical solution for constant cloud properties}

In the case of constant cloud properties ($M_\mathrm{cl}$, $V_\mathrm{cl}$, $A_\mathrm{cl}$, and $C_\mathrm{d}$), the EOM admits analytical solutions expressible in quadrature for arbitrary vertical density and gravitational profiles. 
As shown in Appendix~\ref{app:solution}, the velocity in the dense-cloud limit ($\chi\gg1$) can be written as
\begin{equation}
    \varv_{\mp}(z) = \mp \e^{\pm\int_{z_\mathrm{i}}^{z} \alpha(z') \dif z'} 
    \sqrt{\varv_\mathrm{i}^{2} - 2 \int_{z_\mathrm{i}}^{z} \varg(\tilde{z}) \e^{\mp 2\int_{z_\mathrm{i}}^{\tilde{z}}\alpha(z') \dif z'} \dif \tilde{z}},
    \label{eq:eomsols}
\end{equation}
where $z_\mathrm{i}$ and $\varv_\mathrm{i}$ denote the height and velocity at the beginning of the considered trajectory segment.
The corresponding travel time is given by
\begin{equation}
    \Delta t_{\mp}(z) = \int_{z_\mathrm{i}}^{z} \frac{\mathrm{d}z'}{\varv_{\mp}(z')}.
    \label{eq:tsols}
\end{equation}

The cloud trajectory may consist of one or more segments, each characterized by its own initial conditions $(z_\mathrm{i},\varv_\mathrm{i})$. 
For the first segment these correspond to the launch conditions $(z_0,\varv_0)$. 
The initial direction of motion is determined by the sign of $z_0\varv_0$: if $z_0\varv_0 \le 0$, the cloud moves directly towards the Galactic mid-plane, whereas for $z_0\varv_0 > 0$ it initially moves away from it.
In the latter case the cloud may reach a turnaround height $z_\mathrm{r}$ at which $\varv(z_\mathrm{r}) = 0$ and the motion reverses, after which the subsequent infall constitutes a new trajectory segment with initial conditions $(z_\mathrm{i},\varv_\mathrm{i}) = (z_\mathrm{r},0)$.

Whether such a turnaround occurs or the cloud escapes is determined by the launch velocity relative to the escape velocity. 
In the present model, the symmetry of the background with respect to the Galactic mid-plane (see Section~\ref{sec:bg}) implies that the escape velocity depends only on $|z_0|$.
We therefore define the signed escape velocity as
\begin{equation}
    \varv_\mathrm{esc}(z_0) = \mathrm{sgn}(z_0) \sqrt{2 \int_{|z_0|}^{\infty} \varg(z) \e^{2 \int_{|z_0|}^{z} \alpha(z') \dif z'} \dif z}.
\end{equation}
For $|\varv_0| < |\varv_\mathrm{esc}(z_0)|$ the trajectory turns around at finite height, whereas $|\varv_0| \ge |\varv_\mathrm{esc}(z_0)|$ leads to escape to arbitrarily large distances in the direction of the initial motion.

\subsubsection{Kinematic regimes}

The relation between the initial velocity and the local terminal velocity provides a useful classification of cloud trajectories.
If $|\varv_0| < |\varv_\mathrm{T}(z_0)|$, the cloud accelerates towards the terminal solution, increasing the magnitude of its velocity.
If $|\varv_0| > |\varv_\mathrm{T}(z_0)|$, the cloud initially decelerates, approaching the terminal solution from larger speeds.
Such super-terminal initial conditions may arise transiently, for example through impulsive acceleration during formation or detachment from a larger structure.

The existence of a formal terminal velocity does not imply that the cloud will evolve into a terminal-velocity-dominated state.
The trajectory is instead governed by the global balance between gravity and ram-pressure drag, as characterized by the escape velocity, as well as by the spatial variation of the background medium. 
Therefore, the time-dependence of local conditions may ultimately prevent the cloud from reaching a terminal velocity.

\subsubsection{Scope and limitations}

The analytical solution above applies to constant cloud properties, so that $\alpha$ is a prescribed function of height through the background density $\rho_\mathrm{bg}(z)$.
When time-dependent mass exchange is included, the ram-pressure drag and inertial terms become explicitly coupled to the cloud mass evolution, and the resulting EOM no longer admits analytical solutions expressible in quadrature under realistic Galactic conditions.

The constant-property solution therefore serves as a baseline against which more complex physical processes can be interpreted.
In the following sections we successively relax the assumptions underlying this model and investigate how additional physics modifies the cloud dynamics.

\subsection{Cloud deformation and the Bernoulli effect}
\label{sec:bernoulli}

The baseline model developed above treats the cloud as a coherent object with fixed geometry.
In reality, a cloud moving through the Galactic background medium experiences pressure gradients in the surrounding flow that inevitably lead to deformation.
Such geometric evolution formally enters the EOM through the time dependence of the frontal area $A_\mathrm{cl}(t)$ and drag coefficient $C_\mathrm{d}(t)$, but was neglected in the baseline solution in order to isolate the effects of gravity and ram-pressure drag on the global motion.

In this section, we lift the assumption of fixed geometry and quantify the lateral expansion of the cloud driven by the Bernoulli effect.
To isolate this purely geometric response, we consider deformation at constant cloud mass and neglect simultaneous mass exchange.

\subsubsection{Flow structure and pressure gradients}

\begin{figure*}
    \includegraphics[width=\textwidth]{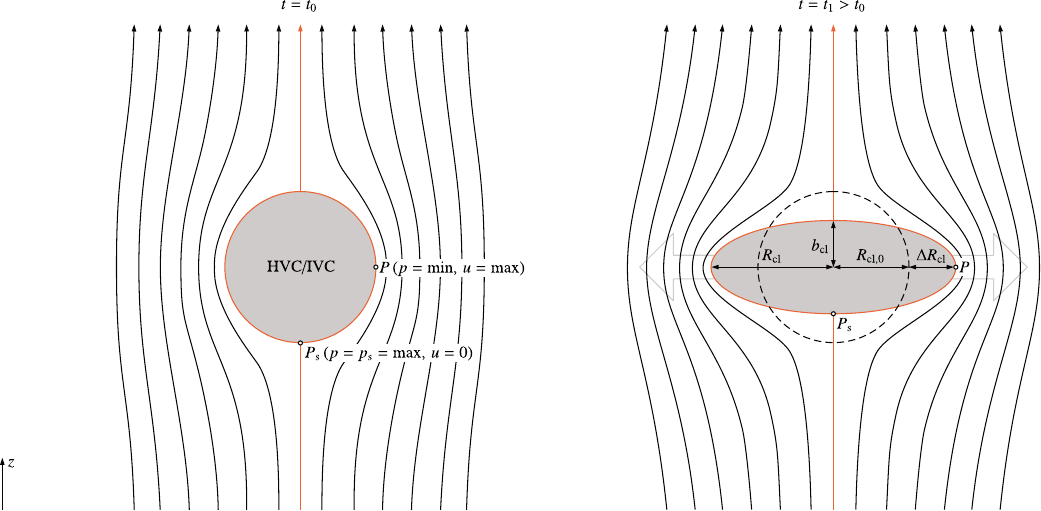}
    \caption{
    Schematic cross-sectional representation of the flow around an initially spherical cloud, shown in the cloud's rest frame during vertical motion towards the Galactic mid-plane. 
    The left panel depicts the undeformed cloud at an early time, indicating the stagnation point at the leading edge~($P_\mathrm{s}$), the body streamline~(red curve), and the lateral pressure minima at the cloud's flanks~(exemplified by the point $P$). 
    The right panel shows the cloud at a later time, after lateral expansion has deformed it into an oblate spheroid, driven by pressure gradients along the body streamline, increasing the effective radius from $R_{\mathrm{cl},0}$ to $R_\mathrm{cl}=R_{\mathrm{cl},0}+\Delta R_\mathrm{cl}$.
    }
    \label{im:bernoulli}
\end{figure*}

A cloud moving vertically through the Galactic background medium develops a stagnation point at its leading edge, where the relative flow velocity vanishes and the pressure reaches a maximum (Fig.~\ref{im:bernoulli}). 
In the rest frame of the cloud, the ambient medium is deflected around the cloud surface, and the streamlines are compressed along its flanks, causing the gas to accelerate through this constricted region.
As a result, the pressure decreases away from the stagnation point and reaches a minimum near the lateral edges of the cloud.

Provided that the adjustment time of the external flow is short compared to the time-scale on which the cloud properties change appreciably, the flow can be treated as quasi-steady.
The high sound speed of the hot halo generally implies subsonic relative motion, while the large Reynolds number confines viscous effects to thin boundary layers.
Under these conditions, the flow around the cloud can be approximated as incompressible and inviscid potential flow.

The corresponding velocity field around a spherical cloud in the dense-cloud limit ($\chi\gg 1$) is derived analytically in Appendix~\ref{app:flowfield}. 
Along the body streamline, the Bernoulli equation of the form
\begin{equation}
    p + \frac{1}{2} \rho_\mathrm{bg} u^2 = p_\mathrm{s}
\end{equation}
applies, where $p_\mathrm{s}$ denotes the stagnation pressure at the leading edge (point $P_\mathrm{s}$ in Fig.~\ref{im:bernoulli}), and $p$ and $u$ are the pressure and flow speed at the lateral point of maximum velocity ($P$). 
Evaluating the velocity field there, namely at the equator of the cloud ($r = R_\mathrm{cl}$, $\theta = \uppi / 2$), yields $u^2 = 9 \varv^2 / 4$ (equation~\ref{eq:sqvelmag}), which implies a pressure difference
\begin{equation}
    \Delta p = p_\mathrm{s} - p = \frac{9}{8} \rho_\mathrm{bg} \varv^2.
    \label{eq:bernoulli_deltap}
\end{equation}

\subsubsection{Lateral expansion of the cloud}

The pressure difference given by equation~\eqref{eq:bernoulli_deltap} drives a net outward force perpendicular to the direction of motion. 
Approximating this force as acting over a characteristic length scale $R_\mathrm{cl}$ yields a radial force density $f_r\simeq\Delta p/R_\mathrm{cl}$. 
Applying Newton's second law to the cloud material then gives the lateral acceleration
\begin{equation}
    \ddot R_\mathrm{cl} = \frac{f_r}{\rho_\mathrm{cl}} \simeq \frac{9 \varv^2}{8 \chi R_\mathrm{cl}},
    \label{eq:rclddot}
\end{equation}
with dots indicating time derivatives.

For sufficiently short times, the deformation remains small ($\Delta R_\mathrm{cl}/R_{\mathrm{cl},0}\ll1$), such that the background conditions and the cloud properties can be treated as locally constant.
In this regime, we approximate $R_\mathrm{cl} \sim R_{\mathrm{cl},0}$ and $\chi \simeq \chi_0$, so that the deceleration parameter can be replaced by its local initial value $\alpha_0 = 3 C_\mathrm{d,0} / (8 \chi_0 R_\mathrm{cl,0})$. 
Hydrodynamic instabilities, which ultimately lead to cloud mass stripping, build up on comparable or longer time-scales (see Section~\ref{sec:insta}) and are therefore neglected at this stage.
The cloud velocity is then well described by the constant-property solution of the EOM in a locally uniform background.
For a cloud falling from $z_0 > 0$ towards the Galactic mid-plane in the dense-cloud limit ($\chi \gg 1$), this solution can be written as
\begin{equation}
    \varv(t) = \varv_\mathrm{T}
    \frac{(\varv_0 + \varv_\mathrm{T}) \e^{\sigma_\mathrm{T} t} + (\varv_0 - \varv_\mathrm{T})}
    {(\varv_0 + \varv_\mathrm{T}) \e^{\sigma_\mathrm{T} t} - (\varv_0 - \varv_\mathrm{T})},
    \label{eq:v_drag_solution}
\end{equation}
with the locally constant terminal velocity $\varv_\mathrm{T} = -\sqrt{\varg_0/\alpha_0}$ and convergence rate $\sigma_\mathrm{T} = 2 \sqrt{\alpha_0 \varg_0}$.
Equivalently, equation~\eqref{eq:v_drag_solution} can be written compactly in terms of a single hyperbolic function. For sub-terminal launches ($|\varv_0|<|\varv_\mathrm{T}|$, including release from rest) it takes the form
\begin{equation}
    \varv(t) = \varv_\mathrm{T} \tanh\!\left[\frac{\sigma_\mathrm{T}}{2} t + \operatorname{artanh}\!\left(\frac{\varv_0}{\varv_\mathrm{T}}\right)\right] ,
    \label{eq:v_drag_tanh}
\end{equation}
whereas for super-terminal launches ($|\varv_0|>|\varv_\mathrm{T}|$) the same expression holds with $\tanh$ and $\operatorname{artanh}$ replaced by $\coth$ and $\operatorname{arcoth}$, reflecting the approach to the terminal value from below or above, respectively. In particular, a cloud released from rest follows the familiar relation $\varv(t)=\varv_\mathrm{T}\tanh(\sigma_\mathrm{T} t/2)$.
Integrating equation~\eqref{eq:rclddot} twice with respect to time yields the effective cloud radius\footnote{Assuming incompressibility, the cloud volume is conserved and the semi-minor axis (see Fig.~\ref{im:bernoulli}) adjusts as $b_\mathrm{cl}(t) \simeq R_{\mathrm{cl},0}^3 / R_\mathrm{cl}^2(t)$.}
\begin{equation}
    R_\mathrm{cl}(t) \simeq R_{\mathrm{cl},0} + \dot R_{\mathrm{cl},0} t 
    + \frac{9}{8 \chi_0 R_{\mathrm{cl},0}} \int_0^t (t-t^\prime) \varv^2(t') \dif t'.
    \label{eq:Rcl_general}
\end{equation}
With $\varv(t)$ given by equation~\eqref{eq:v_drag_solution}, we find
\begin{equation}
    \begin{aligned}
        R_\mathrm{cl}(t) &\simeq R_{\mathrm{cl},0} + \dot R_{\mathrm{cl},0} t \\
        &\quad + \frac{9 \varv_\mathrm{T}}{8 \chi_0 R_{\mathrm{cl},0}}
        \left\{
        \frac{\varv_\mathrm{T}}{2} t^2
        + \frac{2 (\varv_0 + \varv_\mathrm{T})}{\sigma_\mathrm{T}} t\right. \\
        &\left.\qquad\qquad\qquad
        + \frac{4 \varv_\mathrm{T}}{\sigma_\mathrm{T}^2}
        \ln\!\left[
        \frac{2 \varv_\mathrm{T}}{(\varv_0 + \varv_\mathrm{T}) \e^{\sigma_\mathrm{T} t} -(\varv_0 - \varv_\mathrm{T})}
        \right]
        \right\}.
        \end{aligned}
        \label{eq:Rcl_explicit}
\end{equation}

In the ballistic limit ($\alpha_0\rightarrow0$), which provides a good approximation during the early stages of the cloud evolution (see Section~\ref{sec:numresults}), where $\varv(t)=\varv_0-\varg_0 t$, equation~\eqref{eq:Rcl_general} becomes
\begin{equation}
    R_\mathrm{cl}(t) \simeq R_{\mathrm{cl},0}+\dot R_{\mathrm{cl},0}t + \frac{3}{32 \chi_0 R_{\mathrm{cl},0}}
    \left(\varg_0^2 t^4 - 4 \varv_0 \varg_0 t^3 + 6 \varv_0^2 t^2  \right).
\end{equation}
For a cloud initially at rest ($\varv_0 = \dot R_{\mathrm{cl},0} = 0$), the lateral expansion then simplifies to
\begin{equation}
\begin{split}
&\Delta R_\mathrm{cl}(t) \simeq \frac{3 \varg_0^2 t^4}{32 \chi_0 R_{\mathrm{cl},0}} \\
&\quad\simeq \unit[1]{pc}
\left(\frac{\chi_0}{100}\right)^{\!-1}
\left(\frac{R_{\mathrm{cl},0}}{\unit[100]{pc}}\right)^{\!-1}
\left(\frac{\varg_0}{\unit[10^{-8}]{cm\,s^{-2}}}\right)^{\!2}
\left(\frac{t}{\unit[10]{Myr}}\right)^{\!4}.
\end{split}
\end{equation}
This estimate illustrates that even modest vertical velocities can lead to appreciable lateral expansion over Myr time-scales under typical halo conditions.

\subsubsection{Implications for cloud dynamics}

Because the ram-pressure drag force is proportional to the frontal area, Bernoulli-driven lateral expansion feeds back onto the cloud dynamics by enhancing the deceleration. 
As the cloud expands perpendicular to its direction of motion, the effective cross section increases, strengthening the ram-pressure drag and thereby reducing the relative velocity between the cloud and the background medium.
Since the Bernoulli-induced pressure deficit scales with the dynamic pressure of the external flow, $\Delta p \propto \rho_\mathrm{bg} \varv^2$, this feedback simultaneously weakens the driving of further expansion.

Within the idealized coherent and inviscid framework adopted here, drag-induced deceleration alone does not provide a strict saturation mechanism for Bernoulli-driven deformation.
Although drag progressively reduces the cloud velocity, the Bernoulli driving term remains finite as long as the cloud moves relative to the ambient medium.
In the limiting case where the velocity approaches the local terminal value, the dynamic pressure of the flow and therefore the Bernoulli driving term become approximately constant.
Equation~\eqref{eq:rclddot} then continues to predict lateral growth, albeit at a steadily decreasing rate ($\ddot R_\mathrm{cl} \propto R_\mathrm{cl}^{-1}$).
Drag therefore slows the expansion but does not, by itself, halt it within the idealized framework.

In practice, the expansion phase is expected to end once turbulent mixing, fragmentation, and hydrodynamic instabilities dominate the cloud morphology.
In this regime, the concept of a single well-defined cloud radius for a coherent cloud body ceases to be meaningful, and the effective cross section is governed by the evolving ensemble of cloud fragments and turbulent mixing layers.

A word of caution is in order regarding the incompressibility assumption underlying the lateral-expansion model. 
Strictly, the stagnation overpressure at the leading edge relative to the ambient medium, $p_\mathrm{s} - p_\mathrm{bg} \simeq \tfrac{1}{2}\rho_\mathrm{bg}\varv^2$ (from the same Bernoulli relation), compresses the windward face, so the cloud is not perfectly incompressible. 
The magnitude of this effect is set by the Mach number of the cloud's motion through the ambient gas, $\mathrm{Ma}\equiv|\varv|/c_\mathrm{s,bg}$, with ambient sound speed $c_\mathrm{s,bg}=\sqrt{\gamma p_\mathrm{bg}/\rho_\mathrm{bg}}$ and adiabatic index $\gamma=5/3$ for the monatomic gas, through $(p_\mathrm{s}-p_\mathrm{bg})/p_\mathrm{bg}\simeq\tfrac{5}{6}\mathrm{Ma}^2$. 
Because the high sound speed of the hot halo keeps this motion at most transonic ($\mathrm{Ma}\lesssim1$; see Section~\ref{sec:numresults}), the windward compression remains modest, and where the relative flow does steepen into a leading-edge bow shock the strongest compression is confined to a thin post-shock layer rather than restructuring the cloud as a whole.
Moreover, the internal sound-crossing time remains short compared with the deceleration time throughout the coherent phase (Section~\ref{sec:response}), so the interior stays close to pressure equilibrium, and the volume-averaged density that enters $\alpha$
and $\chi$ evolves only gradually until disruption (or, in the cooling runs, condensation) sets in.
The incompressible approximation should therefore be read as a statement about the bulk dynamical response, not about the detailed internal stratification: the leading edge does develop a compressed, quasi-exponential density profile whose axial scale length shortens with increasing deceleration. 
We return to this internal structure as a potential observational diagnostic in Section~\ref{sec:velocity_bridges}.

\subsection{Characteristic time-scales}
\label{sec:timescales}

The EOM introduced in Section~\ref{sec:eom} and its analytical solution presented in Section~\ref{sec:analytics} describe the global vertical motion of a cloud through the Galactic background medium, defining the trajectories permitted by the competition between gravity, ram-pressure drag, and, where applicable, mass exchange.
Whether a cloud actually follows a given trajectory, however, depends on the relative ordering of several characteristic time-scales associated with deceleration, internal adjustment, geometric deformation, instability growth, and thermal evolution.

The purpose of this section is therefore to identify and quantify these time-scales and to place the analytical results derived above into a physical context. 
In particular, they determine whether the kinematic regimes identified in Section~\ref{sec:analytics} can be realized over the cloud lifetime and delineate the conditions under which the assumptions adopted in Section~\ref{sec:bernoulli}, such as quasi-steady external flow and small geometric deformation, remain valid.

For clarity, the main time-scales are grouped into four categories: those governing the global motion of the cloud, those describing the cloud's internal response and deformation, those controlling the growth of hydrodynamic instabilities, and those characterizing its thermal evolution. 
Unless stated otherwise, all time-scales are local quantities that may vary along the cloud trajectory. 
An additional time-scale describing the convergence towards terminal velocity is discussed separately at the end of this section.

\subsubsection{Global motion}
\label{sec:global}

\paragraph*{Free-fall time.}
In the absence of ram-pressure drag and mass exchange ($\alpha=\beta=0$), the cloud follows a purely ballistic trajectory governed solely by the Galactic gravitational field. 
The corresponding free-fall time therefore provides a lower bound on the time required for the cloud to travel between two reference heights, typically the initial height and the Galactic mid-plane, and thus serves as a \emph{global} time-scale for the vertical motion, provided the cloud remains intact.\footnote{This remains true even for $\beta<0$, as mass loss increases the deceleration parameter, $\alpha \propto M_\mathrm{cl}^{-1}$ when the cloud geometry and drag coefficient are held fixed.}

The free-fall time follows from equation~\eqref{eq:tsols}, together with equation~\eqref{eq:eomsols}, by setting $\alpha=0$ and adopting the appropriate integration limits.
For clouds launched from above the Galactic mid-plane ($z_0>0$), this yields
\begin{equation}
\tau_\mathrm{ff} =
\begin{dcases}
\int_{z_0}^{z_\mathrm{r}}
\frac{\mathrm{d}z}{
\sqrt{\varv_0^2 + 2 \int_z^{z_0} \varg(z') \dif z'}
}
\\
\quad + \int_{0}^{z_\mathrm{r}}
\frac{\mathrm{d}z}{
\sqrt{2 \int_z^{z_\mathrm{r}} \varg(z') \dif z'}
},
& 0 < \varv_0 < \varv_\mathrm{esc},
\\
\int_{0}^{z_0}
\frac{\mathrm{d}z}{
\sqrt{\varv_0^2 + 2 \int_z^{z_0} \varg(z') \dif z'}
},
& \varv_0 \le 0,
\end{dcases}
\label{eq:tff}
\end{equation}
which implicitly assumes $\beta=0$ and $\chi\gg 1$.

If the gravitational acceleration varies only weakly over the entire trajectory $[0,z_\mathrm{max}]$, it may be replaced by its path-averaged value
\begin{equation}
    \bar{\varg} = \frac{1}{z_\mathrm{max}} \int_{0}^{z_\mathrm{max}} \varg(z') \dif z'.
\end{equation}
The denominator integrals in equation~\eqref{eq:tff} can then be approximated as $\int_z^{z_*} \varg(z') \dif z' \simeq (z_* - z) \bar{\varg}$, where $z_*\in\{z_0,z_\mathrm{r}\}$, allowing the remaining integrals to be evaluated analytically. 
This yields
\begin{equation}
    \tau_\mathrm{ff} \simeq
    \begin{dcases}
        \frac{1}{\bar{\varg}} \left(\sqrt{2 z_\mathrm{r} \bar{\varg}} - \sqrt{\varv_\mathrm{0}^{2} + 2 (z_0 - z_\mathrm{r}) \bar{\varg}} + |\varv_0|\right), & 0 < \varv_0 < \varv_\mathrm{esc},\\
        \frac{1}{\bar{\varg}} \left(\sqrt{\varv_\mathrm{0}^{2} + 2 z_0 \bar{\varg}} - |\varv_0|\right), & \varv_0 \le 0.
    \end{dcases}
\end{equation}
For clouds starting from rest ($\varv_0 = 0$), we recover the familiar expression $\tau_\mathrm{ff} \simeq \sqrt{2 z_0/\bar{\varg}}$.

\paragraph*{Drag time.}
To quantify how rapidly the cloud momentum is modified by interaction with the background medium, we define a local drag time-scale as the ratio of the cloud momentum to the total retarding force,
\begin{equation}
    \tau_\mathrm{d} = \frac{M_\mathrm{cl} |\varv|}{F_\mathrm{d}},
\end{equation}
where $F_\mathrm{d}$ comprises (i) ram-pressure drag and (ii) an additional `accretion-drag' \citep{Tan:2023} contribution that arises when background material condenses onto the cloud and must be accelerated to the cloud velocity. 
Specifically,
\begin{equation}
    F_\mathrm{d} = F_\mathrm{ram}+F_\mathrm{acc},
\end{equation}
with
\begin{equation}
    F_\mathrm{ram} = \frac{1}{2} C_\mathrm{d} A_\mathrm{cl} \rho_\mathrm{bg} \varv^{2},
    \label{eq:rpdragforce}
\end{equation}
and
\begin{equation}
    F_\mathrm{acc} = \dot{M}_\mathrm{acc} |\varv|,
    \label{eq:facc}
\end{equation}
where $\dot{M}_\mathrm{acc}=\max(\dot{M}_\mathrm{cl},0)$ denotes the net condensation (mass-growth) rate.
This yields the local decomposition
\begin{equation}
    \tau_\mathrm{d} = \left(\frac{1}{\tau_\mathrm{ram}}+\frac{1}{\tau_\mathrm{acc}}\right)^{\!-1},
\end{equation}
i.e.~a harmonic mean of the ram-pressure-drag and accretion-drag time-scales, such that the shorter time-scale dominates.
Here,
\begin{equation}
    \tau_\mathrm{ram} = \frac{2 M_\mathrm{cl}}{C_\mathrm{d} A_\mathrm{cl} \rho_\mathrm{bg} |\varv|} = \frac{1}{\alpha |\varv|},
\end{equation}
and
\begin{equation}
    \tau_\mathrm{acc} = \frac{M_\mathrm{cl}}{\dot{M}_\mathrm{acc}} = \frac{1}{\beta}\quad\text{with $\beta>0$}.
\end{equation}

In adiabatic cases, where net condensation is negligible, $\tau_\mathrm{d} \simeq \tau_\mathrm{ram}$ and deceleration is dominated by ram pressure. 
In cooling-dominated cases, $\tau_\mathrm{acc}$ can become comparable to or shorter than $\tau_\mathrm{ram}$, implying that condensation-driven momentum loading contributes significantly to the effective drag.

\subsubsection{Cloud response and deformation}
\label{sec:response}

\paragraph*{Internal sound-crossing time.}
The sound-crossing time measures how rapidly pressure perturbations propagate across the cloud interior and establish approximate internal pressure equilibrium. 
It is given by
\begin{equation}
    \tau_\mathrm{sc} = \frac{2 R_\mathrm{cl}}{c_\mathrm{s}},
\end{equation}
where $c_\mathrm{s}$ is the internal sound speed of the cloud.

In some of the simulations discussed below, radiative cooling is included but a temperature floor equal to the maximum initial cloud temperature is imposed (see Section~\ref{sec:physics}). 
As a result, the cloud interior cannot cool below its initial temperature and therefore remains approximately isothermal, which is why $c_\mathrm{s}$ is taken to be the isothermal sound speed.

If $\tau_\mathrm{sc}$ is short compared to the global motion time-scales ($\tau_\mathrm{ff}$ or $\tau_\mathrm{d}$), the cloud can adjust quasi-statically to external forcing.
If instead $\tau_\mathrm{sc}$ becomes comparable to or longer than these time-scales, significant internal pressure gradients may develop, invalidating the assumption of a coherent cloud structure.

\paragraph*{Bernoulli (lateral-expansion) time.}
Pressure gradients induced by the Bernoulli effect drive lateral expansion of the cloud, as analysed in Section~\ref{sec:bernoulli}.
A characteristic time-scale for this geometric response can be defined by comparing the cloud radius to its lateral expansion rate,
\begin{equation}
    \tau_\mathrm{B} = \sqrt{\frac{R_\mathrm{cl}}{\ddot R_\mathrm{cl}}}.
\end{equation}
Using equation~\eqref{eq:rclddot}, this yields
\begin{equation}
    \tau_\mathrm{B} \simeq \sqrt{\frac{8 \chi}{9}} \frac{R_\mathrm{cl}}{|\varv|}.
\end{equation}

The condition $\tau_\mathrm{B}\gtrsim\tau_\mathrm{sc}$ ensures that the cloud can adjust internally while deforming, maintaining the validity of the coherent-cloud approximation.
If instead $\tau_\mathrm{B}\lesssim\tau_\mathrm{sc}$, geometric deformation proceeds faster than internal equilibration, producing large-scale distortions of the cloud surface.
The associated increase in effective cross section enhances the ram-pressure drag and can lead to strong deceleration of the cloud, thereby creating favourable conditions for the growth of hydrodynamic instabilities, in particular RT modes
(see below).

\subsubsection{Hydrodynamic instabilities}
\label{sec:insta}

\paragraph*{KH growth time.}
Shear between the moving cloud and the surrounding background medium gives rise to KH instabilities, which strip material from the cloud surface and promote mixing.
For $\chi\gg1$, the KH growth time on scales comparable to the cloud radius is\footnote{This time-scale is comparable to the Bernoulli time (cf.\ Section~\ref{sec:response}), indicating that both processes may become dynamically relevant over similar evolutionary stages.} \citep[e.g.][]{Chandrasekhar:1961,Klein:1994}
\begin{equation}
    \tau_\mathrm{KH} = \sqrt{\chi} \frac{R_\mathrm{cl}}{|\varv|}.
\end{equation}

Although shorter-wavelength modes grow faster, perturbations on scales comparable to $R_\mathrm{cl}$ dominate the global disruption of the cloud and therefore set the relevant instability time-scale.
If $\tau_\mathrm{KH}$ is short compared to the global motion time-scales (Section~\ref{sec:global}), the cloud is efficiently stripped before reaching the Galactic mid-plane.

\paragraph*{RT growth time.}
Whenever the cloud experiences a net deceleration ($\varv \dot{\varv}<0$), RT instabilities develop at its leading edge.
Adopting the cloud radius as the characteristic perturbation scale and assuming a large density contrast, the corresponding growth time may be estimated as
\citep[e.g.][]{Chandrasekhar:1961}
\begin{equation}
    \tau_\mathrm{RT} = \sqrt{\frac{R_\mathrm{cl}}{|\dot{\varv}|}}.
\end{equation}

RT instabilities are therefore most effective during phases of rapid deceleration, such as near turning points of the trajectory or when enhanced drag arises from geometric deformation or mass loading.

\subsubsection{Thermal evolution}
\label{sec:thermal}

\paragraph*{Cooling time.}
The cooling time-scale characterizes how rapidly thermal energy can be removed from gas in the mixing layers formed at the cloud--halo interface.
This gas typically attains intermediate temperatures $T_\mathrm{mix}\sim \sqrt{T_\mathrm{cl} T_\mathrm{bg}}$ \citep{Begelman:1990}, where the cooling function peaks.
The corresponding time is given by
\begin{equation}
    \tau_\mathrm{cool} = \frac{3 k_\mathrm{B} T_\mathrm{mix}}{2 n_\mathrm{mix} \Lambda(T_\mathrm{mix},Z_\mathrm{mix})},
\end{equation}
where $k_\mathrm{B}$ is the Boltzmann constant, $n_\mathrm{mix}$ and $Z_\mathrm{mix}$ are the particle number density and metallicity of the mixed gas, respectively, and $\Lambda(T,Z)$ is the cooling function.

Efficient condensation requires $\tau_\mathrm{cool}\ll\tau_\mathrm{KH}$, such that mixed gas cools and joins the cold phase before being removed by shear-driven instabilities.
If instead $\tau_\mathrm{cool}\gg\tau_\mathrm{KH}$, turbulent mixing dominates and the cloud is gradually eroded into the background medium.

\paragraph*{Evaporation time.}
Thermal conduction transports heat from the hot background into the cold cloud, potentially driving evaporation.
In the classical (unsaturated) conduction regime, the mass-loss rate of an isolated spherical cloud is \citep{Cowie:1977}
\begin{equation}
    \dot{M}_\mathrm{evap} = \frac{16 \uppi \bar{m} \kappa_\mathrm{eff,bg} R_\mathrm{cl}}{25 k_\mathrm{B}},
\end{equation}
where $\kappa_\mathrm{eff,bg}$ is the effective thermal conductivity evaluated at the background temperature $T_\mathrm{bg}$.
The associated evaporation time-scale is then
\begin{equation}
    \tau_\mathrm{evap} = \frac{M_\mathrm{cl}}{\dot{M}_\mathrm{evap}}
    =\frac{25 k_\mathrm{B} \rho_\mathrm{cl} R_\mathrm{cl}^2}{12 \bar{m} \kappa_\mathrm{eff,bg}}.
    \label{eq:tauevap}
\end{equation}

Classical conduction generally provides an upper limit on the cloud lifetime, as strong temperature gradients can lead to saturated heat fluxes that accelerate evaporation.
Conversely, magnetic fields can strongly suppress thermal conduction perpendicular to the field lines, effectively increasing $\tau_\mathrm{evap}$ and enhancing cloud survival.

\paragraph*{Cooling vs.~conduction.}
The competition between radiative cooling and thermal conduction sets a fundamental thermal criterion for cloud survival.
If $\tau_\mathrm{cool}\ll\tau_\mathrm{evap}$, cooling dominates and the cloud can retain or even grow its cold gas reservoir.
If $\tau_\mathrm{evap}\ll\tau_\mathrm{cool}$, conduction overwhelms cooling and the cloud is rapidly dissolved into the ambient halo.

In realistic environments, thermal evolution acts concurrently with hydrodynamic instabilities and geometric deformation.
The ordering of $\tau_\mathrm{cool}$, $\tau_\mathrm{evap}$, and the instability growth times therefore determines whether mass exchange leads to net condensation or disruption, a question that is addressed quantitatively using numerical simulations (see Section~\ref{sec:numresults}).

\subsubsection{Terminal-velocity convergence}
\label{sec:conv}

The time-scales discussed above describe the dynamical, structural, and thermal evolution of the cloud itself.
An additional question is whether the motion approaches the terminal-velocity solution of the EOM when such a solution exists.

Even in a stationary background, the approach to terminal velocity occurs only on a finite time-scale.
Linearizing the EOM (equation~\ref{eq:eom_masschange_compact}) around the instantaneous terminal velocity yields an approximately exponential convergence, $\delta\varv \propto \e^{-t/\tau_\mathrm{T}}$, with convergence time
\begin{equation}
    \tau_\mathrm{T} = \frac{1}{\left| \pm 2 \alpha \varv_\mathrm{T} - \beta \right|}.
\end{equation}

In the absence of mass exchange ($\beta = 0$), this reduces to the classical result $\tau_\mathrm{T} = \left(2 \alpha |\varv_\mathrm{T}|\right)^{\!-1}=\left(2 \sqrt{\alpha |\varg|}\right)^{\!-1}$, where the last equality holds in the dense-cloud limit ($\chi\gg1$).
Mass exchange modifies both the rate of convergence and the stability of the terminal state.
Mass growth ($\beta>0$) increases the effective drag and therefore accelerates convergence, whereas mass loss ($\beta<0$) reduces the effective coupling to the background medium and slows the approach to equilibrium.
If $|\beta|$ approaches $2 \alpha |\varv_\mathrm{T}|$, the convergence time diverges, indicating marginal stability of the terminal solution; for $|\beta| > 2 \alpha |\varv_\mathrm{T}|$ the terminal state becomes formally unstable and cannot be realized dynamically despite its mathematical existence.

A terminal-velocity-dominated regime can be realized only if $\tau_\mathrm{T}$ is short compared to the characteristic time-scales on which the background conditions or cloud properties vary.
In particular, if $\tau_\mathrm{T}$ is comparable to or longer than the global motion time-scales (Section~\ref{sec:global}), the cloud evolves significantly before approaching equilibrium.
More generally, rapid structural or thermodynamic evolution can also preclude terminal behaviour even when a formal solution exists.
The convergence time therefore provides a useful criterion for assessing the applicability of terminal-velocity arguments in a stratified Galactic environment.









\subsection{Mass exchange model}
\label{sec:massex}

Hydrodynamical interactions between a cloud and the surrounding Galactic background medium lead to mass exchange through ablation, turbulent mixing, radiative condensation of ambient gas, and thermal evaporation.
These processes depend on the development of shear layers, hydrodynamic instabilities, and thermal evolution at the cloud--medium interface and therefore cannot be described self-consistently within the one-dimensional dynamical model introduced above.
Instead, we adopt a physically motivated phenomenological prescription that captures the dominant scaling behaviours while remaining compatible with the EOM (equation~\ref{eq:eom_masschange_compact}).

\subsubsection{Instability-driven onset of mass exchange}

In the setups considered here, the cloud mass typically remains close to its initial value during an early phase of coherent motion and changes significantly only after hydrodynamic instabilities have grown to non-linear amplitude.
This behaviour is expected because shear-driven KH instabilities at the cloud--background interface require finite growth times (Section~\ref{sec:insta}) before efficient stripping and mixing can occur.
RT instabilities may become important during phases of strong deceleration but are not required for the onset of mass exchange.

To represent this incubation phase, we introduce a smooth activation function
\begin{equation}
    a(t)=\frac{1}{2} \left[1 + \tanh\!\left(\frac{t - t_M}{\Delta t}\right)\right],
    \label{eq:step}
\end{equation}
which transitions from $a \simeq 0$ at early times to $a \simeq 1$ once instabilities have developed.
The characteristic onset time $t_M$ is therefore expected to be of order the initial KH growth time, $\tau_{\rm KH,0}$, for perturbations on scales comparable to the cloud radius, while $\Delta t$ controls the duration of the transition.
This prescription is sufficiently flexible to also describe cases without a distinct incubation phase, which are recovered in the limit $t_M = 0$ and $\Delta t \to 0$, corresponding to an effectively instantaneous activation.

\subsubsection{Stripping and condensation}

Turbulent mixing across the cloud boundary proceeds on the time-scale set by KH instability growth.
We therefore take the characteristic hydrodynamic mass-exchange rate to scale with the cloud mass divided by the KH growth time,
\begin{equation}
    \dot M_{\mathrm{KH}}(t) = \frac{M_{\mathrm{cl}}(t)}{\tau_{\mathrm{KH}}(t)}.
    \label{eq:mdot_KH}
\end{equation}
For a roughly spherical cloud in the dense-cloud limit, this expression is equivalent to
\begin{equation}
    \dot M_{\mathrm{KH}} = \frac{4}{3} \sqrt{\chi} \rho_{\mathrm{bg}} A_{\mathrm{cl}} |\varv|,
\end{equation}
showing that it corresponds to the incident mass flux across the cloud cross section multiplied by a factor determined by the instability dynamics.

Cooling of mixed gas can reverse the sign of the net mass flux by allowing entrained background material to condense onto the cold phase.
We therefore write the net hydrodynamic contribution as
\begin{equation}
    \dot M_{\mathrm{HD}}(t) = a(t) \dot M_{\mathrm{KH}}(t) \left[\epsilon_{\mathrm{cond}} f_{\mathrm{cool}}(t) - \epsilon_{\mathrm{strip}}\right],
    \label{eq:mdot_hyd}
\end{equation}
where $\epsilon_{\mathrm{strip}}$ and $\epsilon_{\mathrm{cond}}$ are non-negative dimensionless efficiency factors of order unity that control the strength of hydrodynamic stripping and radiative condensation, respectively. 
Setting either parameter to zero suppresses the corresponding process.

Efficient condensation requires that mixed gas cools faster than it is removed by shear-driven instabilities.
Following Section~\ref{sec:thermal}, this condition can be expressed in terms of the ratio of the cooling time to the KH growth time.
We therefore introduce
\begin{equation}
    f_{\mathrm{cool}}(t) = \left\{1 + \left[\frac{\tau_{\mathrm{cool}}(t)}{\eta \tau_{\mathrm{KH}}(t)}\right]^q\right\}^{\!-1},
    \label{eq:fcool}
\end{equation}
where $\eta$ and $q$ are positive constants.
This function smoothly interpolates between inefficient cooling ($f_{\mathrm{cool}}\rightarrow 0$ for $\tau_{\mathrm{cool}}\gg\tau_{\mathrm{KH}}$) and efficient condensation ($f_{\mathrm{cool}}\rightarrow 1$ when $\tau_{\mathrm{cool}}\ll\tau_{\mathrm{KH}}$).

\subsubsection{Thermal evaporation}

Thermal conduction provides an additional channel for mass loss by evaporating cloud material, as discussed in Section~\ref{sec:thermal}.
For the evaporation rate we have
\begin{equation}
    \dot M_{\mathrm{evap}}(t) = \frac{M_{\mathrm{cl}}(t)}{\tau_{\mathrm{evap}}(t)},
\end{equation}
where $\tau_{\mathrm{evap}}$ is the evaporation time-scale defined in Section~\ref{sec:thermal}.
Unlike hydrodynamic stripping, thermal evaporation does not require fully developed instabilities and may operate from the beginning of the evolution. 
Its effectiveness is controlled by an efficiency factor introduced below.

\subsubsection{Total mass evolution}

Combining the above contributions yields the net cloud mass evolution
\begin{equation}
    \dot M_{\mathrm{cl}}(t) = \dot M_{\mathrm{HD}}(t) - \epsilon_{\mathrm{evap}} \dot M_{\mathrm{evap}}(t),
    \label{eq:mdot_total}
\end{equation}
where $\epsilon_{\mathrm{evap}}$ is a non-negative efficiency factor that parametrizes the effectiveness of thermal evaporation.

This prescription naturally reproduces three limiting regimes: (i)~net mass loss when stripping and/or evaporation dominate, (ii)~net mass growth when radiative condensation is efficient, and (iii)~approximately constant mass during the early coherent phase when $a(t)\simeq 0$.
The hydrodynamic contribution retains the fundamental KH-controlled scaling, whereas thermal evaporation represents an independent microphysical process governed by conductive heat transport and characterized by the evaporation time-scale rather than by instability growth.

Given $\dot M_{\mathrm{cl}}(t)$ from equation~\eqref{eq:mdot_total}, the instantaneous cloud mass follows from
\begin{equation}
    M_{\mathrm{cl}}(t) = M_{\mathrm{cl},0} + \int_{0}^{t} \dot M_{\mathrm{cl}}(t') \dif t'.
    \label{eq:massevo}
\end{equation}
The specific mass-exchange rate $\beta(t)$ entering the EOM follows from its definition $\beta=\dot M_{\mathrm{cl}}/M_{\mathrm{cl}}$ (equation~\ref{eq:specmassexrate}). 
Because every contribution to $\dot M_{\mathrm{cl}}$ is proportional to the instantaneous cloud mass, the latter cancels in this ratio and $\beta$ reduces directly to the closed-form expression
\begin{equation}
\begin{split}
    \beta(t) & = \frac{1}{2\tau_{\mathrm{KH}}(t)}\left[1 + \tanh\!\left(\frac{t - t_M}{\Delta t}\right)\right] \\
    &\quad \times \left\{ \frac{\epsilon_{\mathrm{cond}}}{1 + \left[\dfrac{\tau_{\mathrm{cool}}(t)}{\eta \tau_{\mathrm{KH}}(t)}\right]^{q}} - \epsilon_{\mathrm{strip}} \right\} - \frac{\epsilon_{\mathrm{evap}}}{\tau_{\mathrm{evap}}(t)} ,
\end{split}
\label{eq:beta_master}
\end{equation}
which collects the full parameter set $(t_M,\Delta t,\epsilon_{\mathrm{strip}},\epsilon_{\mathrm{cond}},\epsilon_{\mathrm{evap}},\eta,q)$ together with the (in general time-dependent) KH growth, cooling, and evaporation time-scales of Section~\ref{sec:timescales}. 
It makes explicit how the activation function ($t_M$, $\Delta t$), the competition between condensation and stripping ($\epsilon_{\mathrm{cond}}$, $\epsilon_{\mathrm{strip}}$, $\eta$, $q$), and thermal evaporation ($\epsilon_{\mathrm{evap}}$) together set the sign and magnitude of the mass-exchange term. 
Once its parameters are fixed by fitting to the simulated cloud-mass evolution (Section~\ref{sec:numresults}), this is the form evaluated to construct the semi-analytical solutions.

\subsection{Galactic background medium and gravitational potential}
\label{sec:bg}

Evaluation of the cloud dynamics requires specification of the background mass density $\rho_\mathrm{bg}$ and the vertical gravitational acceleration $\varg$ along the cloud trajectory.
Throughout this work, the motion is restricted to the vertical direction at a fixed Galactocentric radius $R$, which is treated as an external parameter.

\begin{figure*}
	\includegraphics[width=\textwidth]{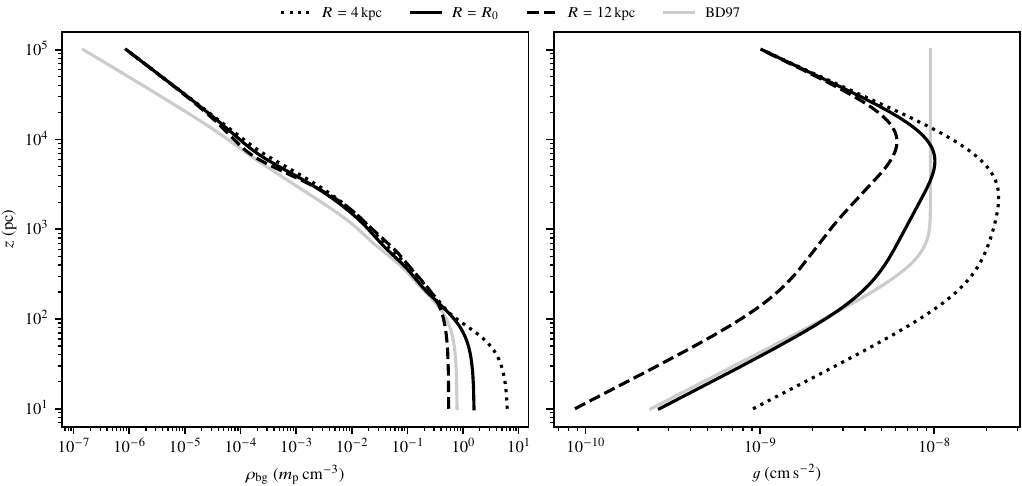}
    \caption{
    Modelled vertical gas density~(left panel) and gravitational acceleration~(right panel) profiles of the Milky Way at Galactocentric radii of $R=\unit[4]{kpc}$~(dotted black lines), $R=R_0=\unit[8.122]{kpc}$~(solid black lines), and $R=\unit[12]{kpc}$~(dashed black lines). 
    These profiles serve as the cloud's background medium throughout this work. For comparison, the grey lines show the solar-neighbourhood profile adopted by \protect\citetalias{Benjamin:1997}. Further details are provided in the text.
    }
    \label{im:bgprofiles}
\end{figure*}

\subsubsection{Background medium}

For the background medium, we adopt the axisymmetric gas density models of \citet{Ferriere:1998} and \citet{Miller:2013}, whose profiles are combined to describe the vertical gas distribution of the Milky Way from the disc to the extended hot halo.
The resulting density profiles are shown in Fig.~\ref{im:bgprofiles} (left panel).

The model of \citet{Ferriere:1998} accounts for the multiphase interstellar medium (ISM), including molecular gas, the cold and warm neutral media (CNM, WNM), the warm ionized medium (WIM), and the hot ionized medium (HIM).
Each phase is characterized by distinct mid-plane densities, radial distributions, and vertical scale heights constrained by a wide range of observational tracers, including CO surveys, \ion{H}{i} 21-cm emission, and pulsar dispersion measures.
We note, however, that this decomposition into discrete ISM phases is necessarily idealized, since the real ISM is dynamic, turbulent, and contains a substantial fraction of gas in thermally unstable regimes \citep[see][]{Heiles:2001,deAvillez:2004,deAvillez:2005}.
At heights beyond approximately 1--\unit[2]{kpc}, we supplement this model with the flattened beta profile of \citet{Miller:2013}, which captures the extended hot halo inferred from \emph{XMM-Newton} \ion{O}{vii} absorption measurements. 

The density profiles exhibit only a weak radial dependence over most of the vertical range. 
While the mid-plane densities are higher at smaller Galactocentric radii, the profiles converge rapidly with increasing height and become nearly indistinguishable in the lower halo. 
Differences remain most pronounced close to the disc and in the transition to the halo component.

For comparison, Fig.~\ref{im:bgprofiles} also includes the most sophisticated solar-neighbourhood profile adopted by \citetalias{Benjamin:1997}, which combines the warm ionized \ion{H}{ii} layer from \citet{Reynolds:1993}, the \ion{H}{i} layers from \citet{Dickey:1990}, and a theoretical isothermal hot halo component from \citet{Wolfire:1995} without radial scaling.

\subsubsection{Gravitational potential}

The vertical gravitational acceleration is derived from the axisymmetric Milky Way potential introduced by \citet{Barros:2016}.
This potential combines several Miyamoto--Nagai disc components \citep{Miyamoto:1975} representing the thin and thick stellar discs and the gaseous disc, together with an analytical bulge model \citep{Hernquist:1990} and a logarithmic dark-matter halo \citep[e.g.][]{Binney:2008}.

The parameters of the individual components are calibrated against the observed Galactic rotation curve as described by \citet{Michtchenko:2023}, providing a self-consistent description of the radial and vertical structure of the Galactic gravitational field.
The resulting vertical acceleration profiles are shown in Fig.~\ref{im:bgprofiles} (right panel), together with the simplified
solar-neighbourhood profile adopted by \citetalias{Benjamin:1997}, based on the disc--halo model of \citet{Wolfire:1995}, for comparison.

In contrast to the density profiles, the vertical gravitational acceleration shows a much stronger radial dependence.
At smaller Galactocentric radii, the magnitude of the acceleration is significantly larger over a wide range of heights, whereas at larger radii it is systematically reduced. 
This separation persists well into the halo and is most pronounced at intermediate heights.

For $|z|>\unit[100]{kpc}$, we impose a quadratic decline of the gravitational acceleration to ensure a physically plausible weakening of the Galactic potential at large distances from the plane.
This modification affects only extreme heights and has a negligible impact on the cloud dynamics in the regime explored here, but becomes relevant when considering trajectories approaching escape conditions.

\subsection{Numerical framework}
\label{sec:num}

To test the dynamical framework developed in the preceding sections and to assess its applicability under more realistic conditions, we perform 3D hydrodynamical simulations using version~3.1 of the Message Passing Interface--Adaptive Mesh Refinement (AMR) Versatile Advection Code \citep[\textsc{mpi--amrvac}\footnote{Available as open source at \url{https://amrvac.org}.};][]{Keppens:2012, Keppens:2021, Keppens:2023, Porth:2014, Xia:2018}.
The simulations solve the full set of compressible hydrodynamical equations on a Cartesian grid.

\subsubsection{Solver setup}
\label{sec:solver}

The code is configured to employ a third-order total variation diminishing (TVD) Runge--Kutta time integration scheme \citep{Shu:1988} with a Courant--Friedrichs--Lewy \citep[CFL;][]{Courant:1928} number of 0.5.
Fluxes are computed using the Harten--Lax--van Leer--Contact (HLLC) approximate Riemann solver \citep{Toro:1994}, combined with a Koren slope limiter \citep{Koren:1993} for third-order spatial reconstruction.
An ideal-gas equation of state with adiabatic index $\gamma=5/3$, appropriate for a monatomic gas and consistent with the value adopted in the analytical treatment (Section~\ref{sec:bernoulli}), is assumed throughout.

\subsubsection{Computational domain and background assumptions}
\label{sec:domain}

The computational domain represents a cuboidal cut-out of the Milky Way at a fixed Galactocentric radius, taken to be the solar radius $R_0$ for all simulations presented here, with dimensions $\unit[1.6]{kpc}\times\unit[1.6]{kpc}\times\unit[6.4]{kpc}$.
A base grid of $32\times32\times128$ cells covers this domain, with the $x$-axis pointing towards the Galactic centre, the $y$-axis aligned with Galactic rotation, and the $z$-axis representing height relative to the Galactic mid-plane.

Leveraging the block-based AMR implementation of \textsc{mpi--amrvac}, we employ blocks of $4^3$ cells at each refinement level and concentrate resolution around steep density and pressure gradients.
In our fiducial setup, five refinement levels are used, yielding a maximum spatial resolution of \unit[3.125]{pc}.
At this highest level, the initial cloud radius is resolved by 32 cells, with each additional grid level increasing the resolution by a factor of two in each spatial dimension.
This places the simulations in the range commonly explored in 3D studies of radiatively cooling cloud evolution, where global quantities such as cloud mass evolution and mixing rates are often found to be substantially more robust than the detailed morphology, although strict convergence can remain difficult to establish and depends on the flow regime, density contrast, and thermal physics \citep[e.g.][]{Cooper:2009,Scannapieco:2015,Schneider:2017,Gronke:2018,Gronke:2022,Abruzzo:2024}.
Convergence of the detailed thermal structure is expected to be more demanding, in particular when the relevant cooling or Field length scales are only marginally resolved \citep{Abruzzo:2024}.
The numerical robustness of the centre-of-mass (COM) motion and mass evolution has been verified by a resolution study (Appendix~\ref{app:resostudy}).

The Galactic mid-plane is located $\unit[1.2]{kpc}$ above the bottom boundary.
No-inflow boundary conditions are applied at both the bottom and top boundaries, while all lateral boundaries are set to be continuous.
Consistent with the analytical framework, we assume that the background density and gravitational acceleration depend only on $z$, neglecting variations along $x$ and $y$ (cf.~Fig.~\ref{im:bgprofiles}).
The background profiles corresponding to the chosen Galactocentric radius are therefore extended uniformly across the horizontal extent of the domain, an assumption justified by the small lateral size of the cloud relative to the box width.

To isolate the cloud dynamics from spurious background motions, the background medium is initialized in hydrostatic equilibrium.
Consequently, the background pressure profile is obtained from
\begin{equation}
    p_\mathrm{bg}(z) = p_\mathrm{bg}(0) - \int_0^{z} \rho_\mathrm{bg}(z') \varg(z') \dif z',
\end{equation}
which is evaluated numerically under the condition that the pressure approaches zero at $|z|=\unit[100]{kpc}$, thereby uniquely fixing the mid-plane pressure $p_\mathrm{bg}(0)$.
The background temperature then follows from the ideal-gas law, $T_\mathrm{bg}(z)=\bar m\,p_\mathrm{bg}(z)/[k_\mathrm{B}\rho_\mathrm{bg}(z)]$, so that the density, gravitational, and hydrostatic-pressure profiles fully determine the vertical thermal structure. 
At the adopted radius $R_0$ this profile passes through $T_\mathrm{bg}\sim\text{few}\times10^{5}\,\mathrm{K}$ near $z\sim\unit[1]{kpc}$, close to the peak of the cooling function (the structure depends on Galactocentric radius through $\rho_\mathrm{bg}$ and $\varg$; cf.~Fig.~\ref{im:bgprofiles}). 
An isolated parcel there would have a cooling time short compared with the cloud's free-fall time and would be thermally unstable. 
A real disc--halo interface, however, does not collapse: feedback processes not modelled here -- supernova and stellar heating, galactic-fountain cycling, and turbulent mixing -- maintain the multiphase hot halo close to a statistical, dynamical equilibrium. 
We therefore treat the background as a controlled proxy for this maintained state, applying radiative cooling and thermal conduction \emph{only} to cloud-tagged gas (region~$\mathcal{C}$; Section~\ref{sec:physics}); this keeps the ambient medium in hydrostatic and thermal equilibrium and isolates the cloud-driven cooling and condensation against a stationary, reproducible reference. 
The idealization it entails -- the absence of the turbulent, multiphase structure and fluctuations of a real hot halo -- is taken up among the limitations (Section~\ref{sec:limitations}).

\subsubsection{Initial conditions and numerical tracers}
\label{sec:cloudinit}

All simulations start from the same initial cloud configuration.
At $t=0$, the cloud is placed at the solar Galactocentric radius $R=R_0$ and at a height $z_0=\unit[5]{kpc}$ above the Galactic mid-plane, corresponding to the Cartesian position $(x_0,y_0,z_0)=(0,0,\unit[5]{kpc})$.
It is initialized as a spherical, homogeneous object composed of purely neutral hydrogen and in pressure equilibrium with the surrounding halo gas.

The initial radius is $R_{\mathrm{cl},0}=\unit[100]{pc}$ and the peak column density is $N_{\ion{H}{i}}=\unit[10^{20}]{cm^{-2}}$.
For $f_\mathrm{cl}=1$ and spherical geometry ($j=1$), the resulting initial mass follows from equation~\eqref{eq:initcloudmass}, $M_{\mathrm{cl},0}\simeq \unit[1.7\times10^{4}]{M_{\sun}}$.
The cloud is launched with an initial vertical velocity $\varv_0 = -\unit[100]{km\,s^{-1}}$ and zero transverse components.
This corresponds to an initial KH growth time $\tau_\mathrm{KH,0} = \sqrt{\chi_0} R_{\mathrm{cl},0} / |\varv_0| \simeq \unit[17.8]{Myr}$ (for $\chi_0 \simeq 331$), which we adopt as the reference time-scale for normalizing the onset time and transition width of the mass-exchange model (see Table~\ref{tab:mexparams}).
For comparison, the corresponding free-fall time is $\tau_\mathrm{ff}\simeq\unit[32.9]{Myr}$.

In the numerical implementation, the cloud occupies the finest grid cells and is tagged by a passive scalar $C$ with unit concentration, while the ambient medium is assigned zero tracer abundance.
This scalar is advected with the flow and allows cloud material to be tracked throughout the evolution, including after strong deformation and mixing with the background gas.

\subsubsection{Region definitions}
\label{sec:regions}

At each time $t$, several regions are defined based on tracer abundance and gas density.

\begin{itemize}
\item Cells with tracer concentration $C \ge 10^{-6}$ are classified as cloud-related material.
The set of all such cells defines the cloud-related region~$\mathcal{C}(t)$.

\item Within $\mathcal{C}(t)$, the dense bulk cloud is identified by a density threshold.
Cells with densities exceeding one third of the initial cloud density $\rho_{\mathrm{cl},0}$ define the bulk region~$\mathcal{B}(t)$, representing the coherent cloud body.

\item Cloud-related cells with lower densities constitute the diffuse cloud-related component, $\mathcal{D}(t)\equiv\mathcal{C}(t)\setminus\mathcal{B}(t)$.
This region includes stripped gas, turbulent wakes, and cloud material that has been diluted by interaction with the ambient halo.

\item To characterize actively mixing gas, we further define a mixing region~$\mathcal{M}(t)\subset\mathcal{D}(t)$ consisting of cells with tracer values $C\ge 10^{-2}$ and densities in the range $0.01 \rho_{\mathrm{cl},0} \le \rho < \rho_{\mathrm{cl},0}/3$.
This criterion selects gas that still originates predominantly from the cloud but has been substantially mixed and diluted by the background medium.
\end{itemize}

Such threshold-based classifications are commonly employed in numerical studies of cloud--halo interactions, and our results were verified to be insensitive to moderate variations of the adopted thresholds.

\subsubsection{Kinematic and drag diagnostics}
\label{sec:drag}

For any quantity $Q$, the mass-weighted average over an arbitrary region~$\Omega(t)$ is defined as
\begin{equation}
    \langle Q \rangle_{\Omega}(t) = \frac{1}{M_{\Omega}(t)} \int_{\Omega(t)} \rho(\mathbfit{x},t) Q(\mathbfit{x},t) \dif V,
\end{equation}
where
\begin{equation}
    M_{\Omega}(t) = \int_{\Omega(t)} \rho(\mathbfit{x},t) \dif V
\end{equation}
is the gas mass contained in $\Omega$, with $\rho$ denoting the local gas density.
In the following, $\Omega$ refers to one of the regions introduced in Section~\ref{sec:regions}, namely the bulk cloud~$\mathcal{B}(t)$, the diffuse cloud-related component~$\mathcal{D}(t)$, or the mixing region~$\mathcal{M}(t)$, with corresponding masses
\begin{equation}
    M_\mathrm{cl} \equiv M_{\mathcal{B}},\qquad
    M_\mathrm{diff} \equiv M_{\mathcal{D}},\qquad
    M_\mathrm{mix} \equiv M_{\mathcal{M}}.
\end{equation}

Using the bulk region, the instantaneous COM position and velocity of the cloud are
\begin{equation}
    \mathbfit{r}(t) = \langle \mathbfit{x} \rangle_{\mathcal{B}},\qquad
    \boldsymbol{\varv}(t) = \langle \mathbfit{u} \rangle_{\mathcal{B}},
\end{equation}
where $\mathbfit{u}=(u_x,u_y,u_z)^{\mathsf{T}}$ denotes the local flow velocity; $\varv(t)$ refers to the $z$-component of $\boldsymbol{\varv}(t)$.

Mass-weighted velocity dispersions are used as diagnostics of turbulent and non-coherent motions and are defined for any region~$\Omega$ as
\begin{equation}
    \sigma_{\Omega}(t) = \sqrt{\left\langle\left|\mathbfit{u} - \langle\mathbfit{u}\rangle_{\Omega}\right|^{2}\right\rangle_{\Omega}}.
\end{equation}
We evaluate $\sigma_{\mathrm{cl}}$, $\sigma_{\mathrm{diff}}$, and $\sigma_{\mathrm{mix}}$ for the bulk, diffuse, and mixing regions, respectively.

We estimate the cloud's effective frontal area $A_\mathrm{cl}(t)$ as the projected area of the bulk-cloud material perpendicular to the direction of motion. 
Numerically, this is obtained by projecting all bulk cells onto the finest Cartesian $xy$ grid and summing the corresponding cell-face areas. 
The corresponding area-equivalent radius is then
\begin{equation}
    R_\mathrm{cl}(t)=\sqrt{A_\mathrm{cl}(t)/\uppi} .
\end{equation}

Rearranging the EOM (equation~\ref{eq:eom_masschange}) allows the instantaneous ram-pressure drag force acting on the cloud to be inferred from the cloud acceleration and mass evolution along its trajectory,
\begin{equation}
    F_\mathrm{ram}(t) = -\mathrm{sgn}(\varv)
    \Big[
    M_\mathrm{cl} \dot{\varv}
    +
    \dot{M}_\mathrm{cl} \varv
    +
    \big(M_\mathrm{cl} - \rho_\mathrm{bg} V_\mathrm{cl}\big) \varg
    \Big],
\end{equation}
where all cloud quantities refer to the bulk region~$\mathcal{B}(t)$, and the background properties are evaluated at the COM height of the cloud.
For physical ram-pressure drag, $F_\mathrm{ram}(t)$ is non-negative for motion in either direction, since drag always opposes the cloud velocity.
In practice, small negative values may arise from numerical noise in the time derivatives or from deviations between the bulk-cloud description and the one-dimensional EOM; such values are therefore not interpreted as physical drag.
The drag coefficient then follows from the standard relation (cf.~equation~\ref{eq:rpdragforce}),
\begin{equation}
    C_\mathrm{d}(t) = \frac{2 F_\mathrm{ram}(t)}{\rho_\mathrm{bg}(t) \varv^{2}(t) A_\mathrm{cl}(t)} .
\end{equation}

\subsubsection{Simulation types and applied physics}
\label{sec:physics}

We perform three types of simulations: purely adiabatic runs, runs with radiative cooling only (hereafter RC), and runs with both radiative cooling and thermal conduction (hereafter RC+TC).
To isolate the thermal evolution of cloud material, cooling and conduction are applied only within region~$\mathcal{C}(t)$.
This restriction ensures that the background medium remains in both hydrostatic and thermal equilibrium throughout the simulation.

To prevent premature cloud collapse, we impose a temperature floor below which cooling is disabled, set equal to the maximum initial cloud temperature.
Radiative cooling follows the collisional ionization equilibrium cooling curve of \citet{Sutherland:1993} for gas with metallicity $[\mathrm{Fe}/\mathrm{H}]=-0.5$ (or, equivalently, $Z\simeq \unit[0.3]{Z_{\sun}}$), consistent with halo metallicity estimates \citep{Miller:2015}.

Thermal conduction is modelled using the classical Spitzer conductivity, appropriate for high-temperature, low-density plasmas in which electron collisions dominate \citep{Spitzer:1962}.
The conductive heat flux is written as
\begin{equation}
    \mathbfit{q} = -\kappa_\mathrm{eff} \nabla T ,
\end{equation}
with an effective conductivity
\begin{equation}
    \kappa_\mathrm{eff} = f_\mathrm{sup} \kappa_\mathrm{Sp} ,
\end{equation}
where
\begin{equation}
    \kappa_\mathrm{Sp} = \frac{1.84 \times 10^{-5}}{\ln \Psi}
    \left(\frac{T}{\mathrm{K}}\right)^{\!5/2}
    \,\mathrm{erg\,s^{-1}\,cm^{-1}\,K^{-1}}
\end{equation}
is the classical Spitzer conductivity.
Here, $T$ denotes the gas temperature, and the Coulomb logarithm is fixed to $\ln \Psi = 35$, appropriate for typical electron densities and temperatures in the Galactic halo \citep{Huba:2016}.

Magnetic suppression of thermal conduction is approximated by adopting an efficiency factor $f_\mathrm{sup}=0.1$.
This choice is motivated by the results of \citet{Kooij:2021}, who showed that magnetic fields draped around cold clouds in a circumgalactic medium suppress heat transport perpendicular to the field lines, reducing the conductive heat flux to approximately 3--\unit[15]{per cent} of the Spitzer value.

To avoid unphysical heat fluxes in regions with steep temperature gradients, we apply the saturation prescription of \citet{Cowie:1977},
\begin{equation}
    q_\mathrm{sat} = 5 \phi_\mathrm{sat} \rho c_\mathrm{s}^3 ,
\end{equation}
with $\phi_\mathrm{sat} = \sqrt{f_\mathrm{sup}}$ following \citet{Armillotta:2017}.

Conduction is solved in an operator-split manner using an explicit super-time-stepping integrator with second-order accuracy in time, while a slope-limited symmetric scheme is used for stable and accurate spatial discretization \citep{Xia:2018}.

\section{Results}
\label{sec:results}
In this section we present the implications of the dynamical framework developed in Section~\ref{sec:dynamics} and compare them with the outcomes of 3D hydrodynamical simulations.
We begin with the analytically tractable constant-property limit, in which the cloud mass, geometry, and drag coefficient are held fixed.
After discussing clouds released from rest, we examine the effect of non-zero initial velocity.
We then turn to the hydrodynamical simulations for the adiabatic, RC, and RC+TC setups, assessing how mass exchange and cloud deformation modify the trajectories and to what extent the semi-analytical model remains applicable.

Unless stated otherwise, the analytical and semi-analytical results presented here assume spherical cloud geometry ($j=1$), a neutral fraction $f_\mathrm{cl}=1$, a large density contrast ($\chi \gg 1$), and a constant drag coefficient $C_\mathrm{d}=1$, representative of a blunt body in a high-Reynolds-number flow.
The analytical solutions correspond to the constant-property limit without mass exchange, whereas the semi-analytical solutions additionally include mass-loss and mass-growth terms as described in Section~\ref{sec:massex}.

\subsection{Analytical results for constant-property clouds}
\label{sec:anaresults}
\subsubsection{Clouds released from rest}
\label{sec:ana_v0zero}

\begin{figure*}
	\includegraphics[width=\textwidth]{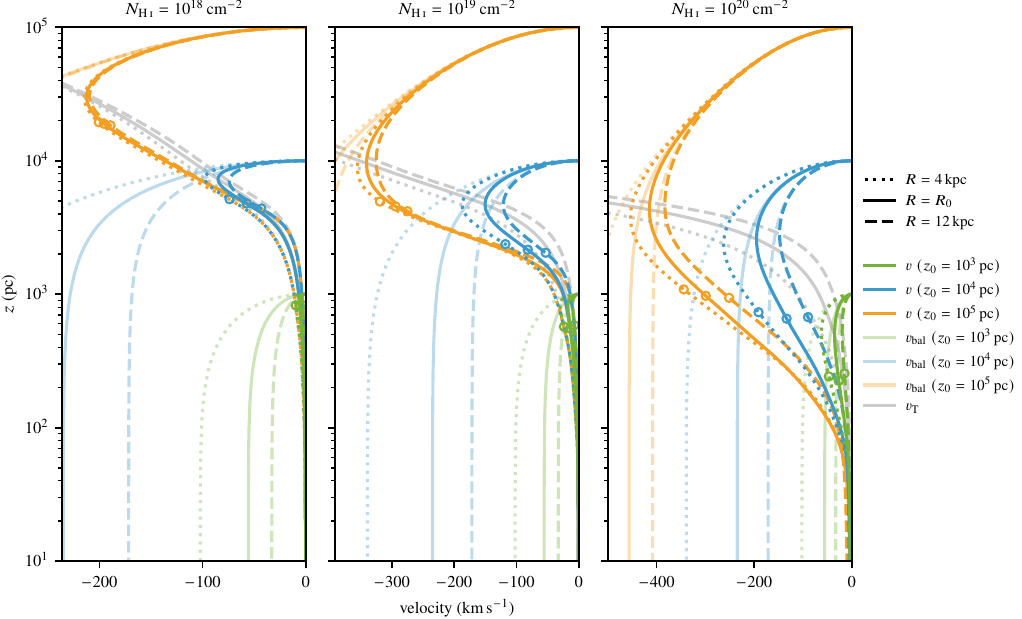}
    \caption{
    Trajectories of constant-property clouds (opaque coloured curves) with peak column densities $N_{\ion{H}{i}}=\unit[10^{18}]{cm^{-2}}$~(left panel), $\unit[10^{19}]{cm^{-2}}$~(middle panel), and $\unit[10^{20}]{cm^{-2}}$~(right panel), released from rest at heights $z_0= \unit[10^3]{pc}$~(green), $\unit[10^4]{pc}$~(blue), and $\unit[10^5]{pc}$ ~(orange), and at Galactocentric radii $R=\unit[4]{kpc}$~(dotted), $R_0=\unit[8.122]{kpc}$~(solid), and $\unit[12]{kpc}$~(dashed). 
    Empty circles mark the cloud positions after one free-fall time. 
    Ballistic trajectories (semi-transparent coloured curves) and local terminal-velocity curves (grey curves) are shown for comparison.
    }
    \label{im:trajv0zero}
\end{figure*}

Figure~\ref{im:trajv0zero} shows the trajectories of clouds released from rest over a wide range of column densities, initial heights, and Galactocentric radii.

All trajectories exhibit a similar qualitative evolution.
Starting from rest, the clouds initially accelerate towards the Galactic mid-plane in a nearly ballistic phase during which ram-pressure drag is dynamically negligible and the motion closely follows the ballistic reference solutions (semi-transparent coloured curves in Fig.~\ref{im:trajv0zero}).
As the clouds descend into denser regions of the halo, the drag force increases and eventually becomes comparable to gravity, causing the infall speed to reach a maximum.
The motion then evolves towards the local terminal-velocity curves (grey curves).

\begin{figure*}
	\includegraphics[width=\textwidth]{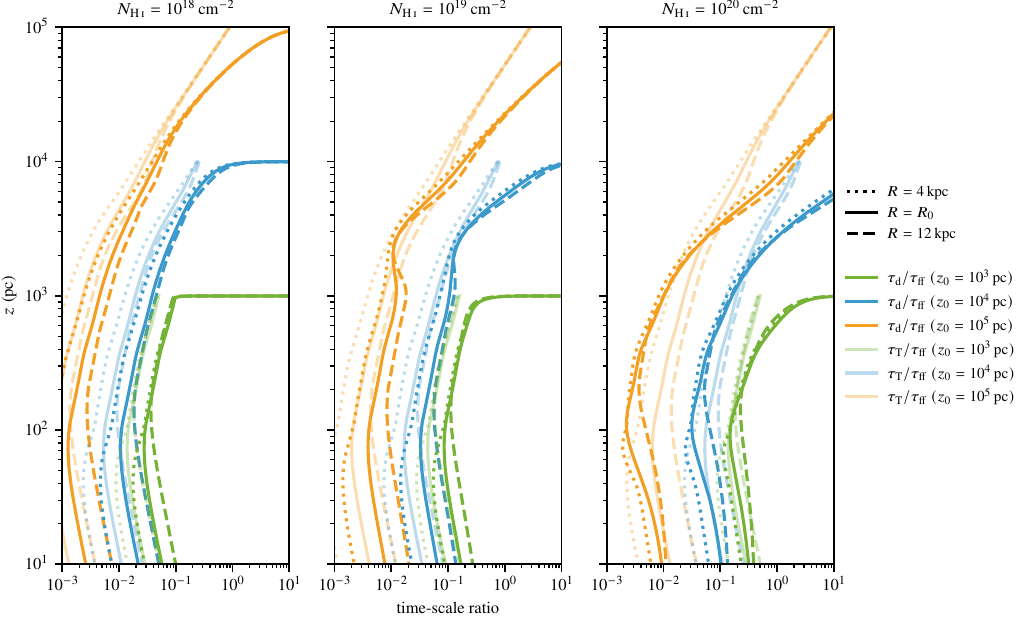}
    \caption{
    Positional evolution of the drag time (opaque curves) and the terminal-velocity convergence time (semi-transparent curves), both normalized to the free-fall time, for constant-property clouds with peak column densities $N_{\ion{H}{i}}=\unit[10^{18}]{cm^{-2}}$~(left panel), $\unit[10^{19}]{cm^{-2}}$~(middle panel), and $\unit[10^{20}]{cm^{-2}}$~(right panel), released from rest at heights $z_0= \unit[10^3]{pc}$~(green), $\unit[10^4]{pc}$~(blue), and $\unit[10^5]{pc}$ ~(orange), and at Galactocentric radii $R=\unit[4]{kpc}$~(dotted), $R_0=\unit[8.122]{kpc}$~(solid), and $\unit[12]{kpc}$~(dashed).
    }
    \label{im:timesv0zero}
\end{figure*}

The dynamical origin of this behaviour is illustrated in Fig.~\ref{im:timesv0zero}, which shows the positional evolution of the drag time $\tau_\mathrm{d}$ and the terminal-velocity convergence time $\tau_\mathrm{T}$, both normalized to the free-fall time.
At the moment of release the drag time formally diverges because the cloud velocity is zero.
As the cloud begins to accelerate under gravity, however, $\tau_\mathrm{d}$ rapidly decreases from this formally infinite value, dropping by several orders of magnitude while the cloud has moved only a small distance.

During this initial adjustment, the ratio $\tau_\mathrm{d}/\tau_\mathrm{ff}$ may already approach or even fall below unity for some parameter combinations.
Farther along the trajectory, its variation becomes more gradual as the cloud continues its descent through the halo and the increasing background density strengthens the drag force.

At the same time the convergence time $\tau_\mathrm{T}$ becomes much shorter than the free-fall time, implying that the velocity adjusts rapidly to the local terminal solution.
Three characteristic dynamical regimes can therefore be identified: an initial ballistic phase ($\tau_\mathrm{d}\gg\tau_\mathrm{ff}$), a drag-influenced regime in which $\tau_\mathrm{d}$ becomes comparable to $\tau_\mathrm{ff}$, and a terminal regime in which the velocity rapidly approaches the local terminal value ($\tau_\mathrm{T}\ll\tau_\mathrm{ff}$).

The detailed behaviour depends strongly on the cloud properties and environment.
Low-column-density clouds rapidly converge towards the terminal velocity, whereas clouds with larger column densities possess greater inertia per unit area and therefore remain far from terminal equilibrium over extended distances.
This trend is also reflected in the characteristic time-scales shown in Fig.~\ref{im:timesv0zero}: both $\tau_\mathrm{d}$ and $\tau_\mathrm{T}$ shift systematically towards larger values with increasing column density, reflecting the weaker dynamical coupling of more massive clouds to the ambient medium.

The initial height $z_0$ controls the available acceleration time.
Clouds released from larger heights traverse extended regions of low ambient density and therefore experience relatively weak drag during the early stages of their descent.
Their motion consequently remains close to the ballistic solution over a larger range of heights.
Only once the clouds reach denser regions of the halo does ram-pressure drag become dynamically important and the trajectories begin to deviate significantly from the ballistic curves.
Nevertheless, because the ram-pressure drag force scales as $F_\mathrm{ram} \propto \rho_\mathrm{bg} \varv^2$, the rapidly increasing infall speed can partially compensate for the low background density at large altitude, so that deviations from purely ballistic motion may already become noticeable before the clouds reach the densest regions of the halo.

The Galactocentric radius introduces an additional dependence through the radial variation of both the gravitational field and the ambient density structure.
At smaller radii, stronger gravity and higher ambient densities shorten the dynamical time-scales, whereas weaker forces at larger radii allow quasi-ballistic motion to persist over longer distances.

Taken together, these dependencies imply that the transition from gravity-dominated to drag-dominated motion occurs earliest for clouds with low column density, small initial height, and large Galactocentric radius.
In contrast, dense clouds released from large heights at small radii can remain in a quasi-ballistic regime over much of their trajectories.
This behaviour is consistent with the analysis of \citetalias{Benjamin:1997}, who showed that sufficiently massive halo clouds need not follow terminal-velocity motion but may instead experience extended phases of near-ballistic infall.
The present results extend that picture by demonstrating the systematic dependence of this transition on Galactocentric radius in a stratified background medium.

The peak speeds reached in some parameter combinations can become very large, particularly for high column densities and large initial heights.
These values reflect the idealized assumption of constant cloud properties adopted in the analytical model and illustrate the dynamical response of massive clouds in a stratified Galactic potential prior to significant mass loss or deformation, rather than representing typical observed halo clouds.

\subsubsection{Non-zero initial velocities}
\label{sec:ana_v0nonzero}

\begin{figure*}
	\includegraphics[width=\textwidth]{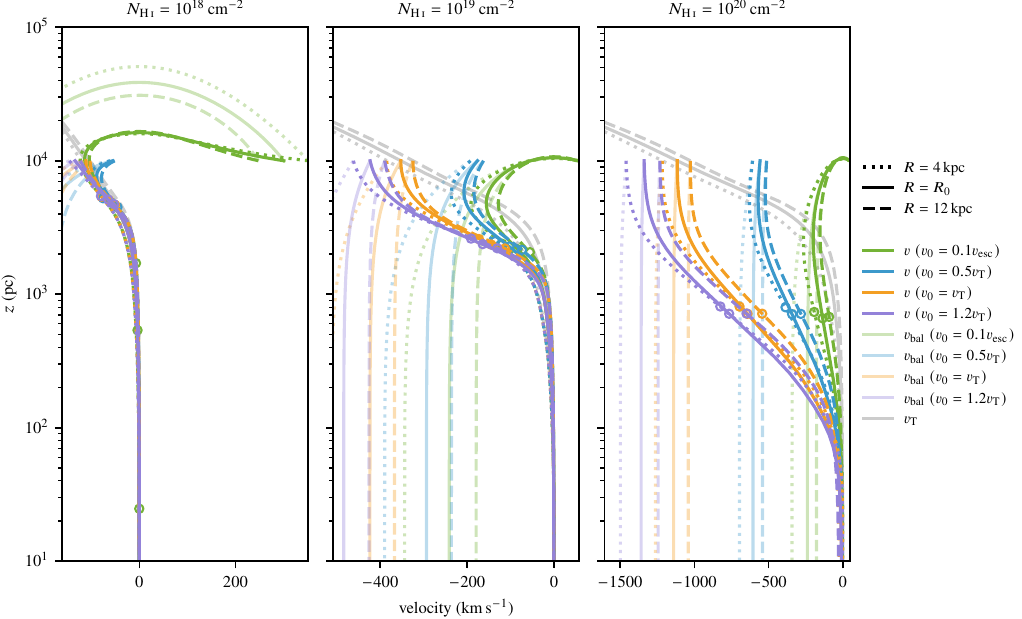}
    \caption{
    As for Fig.~\ref{im:trajv0zero}, but for clouds launched from a common height $z_0=\unit[10^4]{pc}$ with non-zero initial velocities: $\varv_0=0.1 \varv_\mathrm{esc}$~(green), $0.5 \varv_\mathrm{T}$~(blue), $\varv_\mathrm{T}$~(orange), and $1.2 \varv_\mathrm{T}$~(purple).
    }
    \label{im:trajv0nonzero}
\end{figure*}

Unlike the previous subsection, which explored how cloud properties and environment determine the trajectories of clouds released from rest, we now isolate the effect of non-zero initial velocities by launching clouds from a common height with different initial speeds (Fig.~\ref{im:trajv0nonzero}).

At early times the motion naturally reflects the imposed initial velocity.
Clouds launched away from the Galactic mid-plane (green curves) first decelerate under gravity, eventually reverse direction, and subsequently evolve towards the mid-plane in a manner similar to the zero-velocity case.
These outflow cases confirm that the closed-form solution applies to launches directed away from the plane, including the turnaround, and not only to direct infall.
Clouds launched towards the Galactic mid-plane initially accelerate or decelerate depending on whether the initial speed is below or above the local terminal speed.

Sub-terminal clouds ($|\varv_0|<|\varv_\mathrm{T}|$; blue curves) initially accelerate because gravity exceeds the ram-pressure drag force, whereas super-terminal clouds ($|\varv_0|>|\varv_\mathrm{T}|$; purple curves) experience an initial phase of drag-dominated deceleration.
In both cases the velocity subsequently evolves towards the height-dependent terminal solution.

\begin{figure*}
	\includegraphics[width=\textwidth]{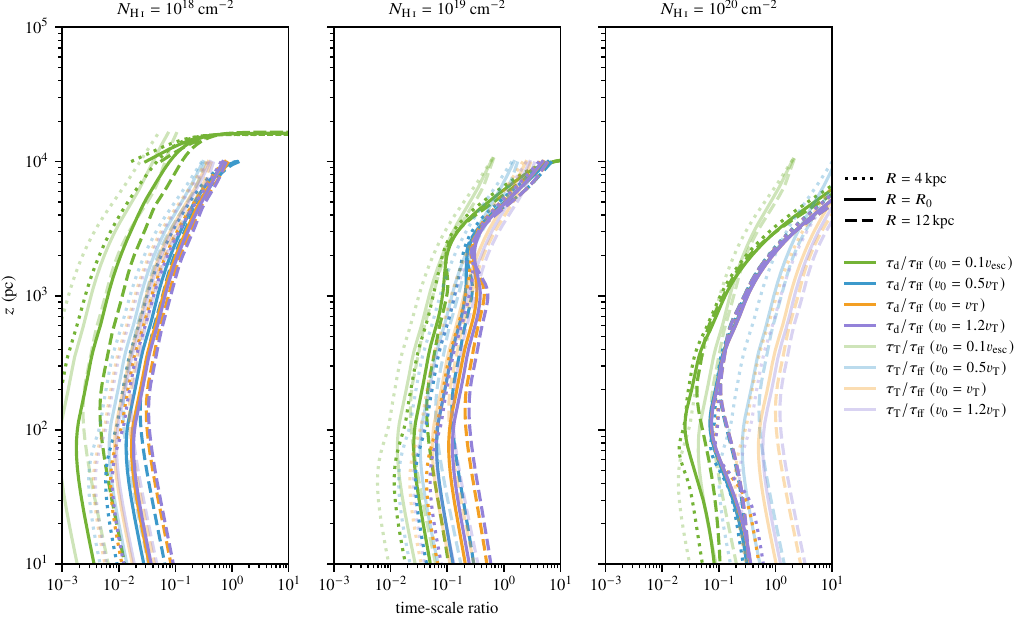}
    \caption{
    As for Fig.~\ref{im:timesv0zero}, but for clouds launched from a common height $z_0=\unit[10^4]{pc}$ with non-zero initial velocities: $\varv_0=0.1 \varv_\mathrm{esc}$~(green), $0.5 \varv_\mathrm{T}$~(blue), $\varv_\mathrm{T}$~(orange), and $1.2 \varv_\mathrm{T}$~(purple).
    }
    \label{im:timesv0nonzero}
\end{figure*}

The relaxation behaviour is illustrated in Fig.~\ref{im:timesv0nonzero}, which shows the positional evolution of the drag time $\tau_\mathrm{d}$ and the convergence time $\tau_\mathrm{T}$.
Because the initial velocity is non-zero in this case, the drag time is finite at the starting height.
As the cloud begins to accelerate or decelerate away from its initial velocity, $\tau_\mathrm{d}$ changes rapidly while the cloud has moved only a small distance, after which the variation becomes more gradual as the cloud continues its descent through the halo.
The convergence time $\tau_\mathrm{T}$ likewise decreases along the trajectory and in many cases eventually becomes much shorter than the free-fall time, at which point the velocity adjusts rapidly to the local terminal solution (grey curves in Fig.~\ref{im:trajv0nonzero}).

Even clouds launched exactly at the local terminal velocity do not remain in equilibrium: because both gravity and ambient density vary with height, the terminal velocity itself evolves along the trajectory.
A cloud that initially satisfies $\varv_0=\varv_\mathrm{T}(z_0)$ therefore immediately departs from equilibrium as it moves into regions with different background conditions.

Although the initial velocity can significantly modify the early trajectory, the long-term evolution is largely governed by the same gravity--drag balance that controls the zero-velocity solutions.
The terminal velocity therefore acts as a dynamical attractor at late times, although substantial departures from equilibrium can persist over large distances for clouds with large inertia per unit area or strongly non-equilibrium initial conditions.

These analytical results provide a reference framework against which the outcomes of the full hydrodynamical simulations are evaluated in the following section.

\subsection{Hydrodynamical simulations}
\label{sec:numresults}

All simulations adopt the same initial cloud: a peak neutral-hydrogen column density $N_{\ion{H}{i}}=\unit[10^{20}]{cm^{-2}}$ and radius $R_{\mathrm{cl},0}=\unit[100]{pc}$ (Section~\ref{sec:cloudinit}). 
This places the run at the upper end of the observed HVC range: the \ion{H}{i} column densities of HVCs span roughly $10^{18}$--$\unit[10^{20}]{cm^{-2}}$, with the densest cores of complexes such as Complex~C reaching $N_{\ion{H}{i}}\sim\unit[10^{20}]{cm^{-2}}$ \citep[e.g.][]{Wakker:1997,Putman:2012}; the chosen value is thus representative of the densest neutral clouds observed in the Milky Way halo. 
It is also the most demanding case for the terminal-velocity paradigm, since such high-inertia clouds couple weakly to the ambient medium and converge to terminal motion most slowly (Section~\ref{sec:anaresults}). 
The complementary low-column-density regime is treated analytically in Section~\ref{sec:anaresults} and left to future numerical work (Section~\ref{sec:limitations}).

Figures~\ref{im:timeseries_adiabatic}--\ref{im:timeseries_rctc} present time series of density and temperature slices for the adiabatic, RC, and RC+TC runs, illustrating the morphological evolution of the cloud and its interaction with the Galactic background medium. 
Animations of the time evolution are available as supplementary online material.

The same evolution is summarized in Fig.~\ref{im:comparison}, which shows the temporal behaviour of key dynamical and thermodynamic cloud properties.
Opaque coloured curves represent the simulations, while semi-transparent curves of the same colour show the corresponding semi-analytical solutions, and grey curves indicate analytical reference solutions where available.

\begin{figure}
    \includegraphics[width=\columnwidth]{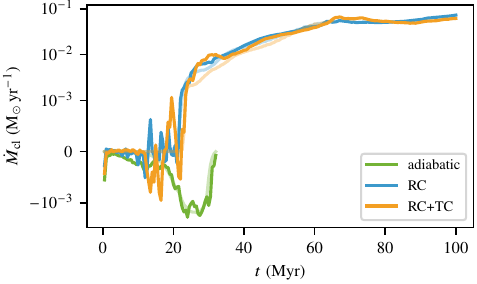}
    \caption{
    Simulated bulk-cloud mass-exchange rates (opaque coloured curves) together with their least-squares fits (semi-transparent coloured curves), based on the model described in Section~\ref{sec:massex}.
    The corresponding fit parameters are listed in Table~\ref{tab:mexparams}.
    }
    \label{im:massexfit}
\end{figure}

The parameters of the phenomenological mass-exchange model introduced in Section~\ref{sec:massex} are determined by least-squares fits to the simulated cloud-mass evolution.
Figure~\ref{im:massexfit} compares the simulated mass-exchange rates with the fitted model, and the resulting parameter values are summarized in Table~\ref{tab:mexparams}.
These parameters are then used to construct the semi-analytical solutions shown in Fig.~\ref{im:comparison}.

For reference, we also compare the simulated cloud radius with the analytical Bernoulli expansion model derived in Section~\ref{sec:bernoulli}. 
The corresponding solution (shown in Fig.~\ref{im:comparison}c) describes the purely geometric response of the cloud to the Bernoulli pressure gradient and is not included in the analytical or semi-analytical trajectory models.
The associated Bernoulli time-scale is not shown explicitly in Fig.~\ref{im:comparison}h, as it closely tracks the KH growth time (see Section~\ref{sec:timescales}) and is therefore effectively captured by the latter.

A robust feature of Fig.~\ref{im:comparison}h, common to all three runs, is that once the initial transient has decayed the dynamical time-scales -- drag, convergence, sound-crossing, and KH/RT growth -- collapse into a narrow band of order unity, while the two thermal time-scales separate out: the cooling time drops well below this band and the evaporation time runs off above it. 
The clustering is physical rather than incidental: these dynamical times all reduce to the same algebraic combination of $\chi$, $R_\mathrm{cl}$, and $|\varv|$, which adjust together as the cloud is compressed and decelerated, so that deceleration, deformation, and disruption proceed simultaneously on roughly the dynamical time.
The thermal times, by contrast, are governed by the strongly non-linear cooling (and conduction) functions and therefore scale quite differently: a cooling time well below $\tau_\mathrm{ff}$ keeps condensation active, whereas the evaporation time grows by orders of magnitude and becomes dynamically negligible. 
This is precisely the marginally coupled regime in which the constant-property terminal-velocity description is least reliable.

\subsubsection{Adiabatic run}
\begin{figure*}
	\includegraphics[width=\textwidth]{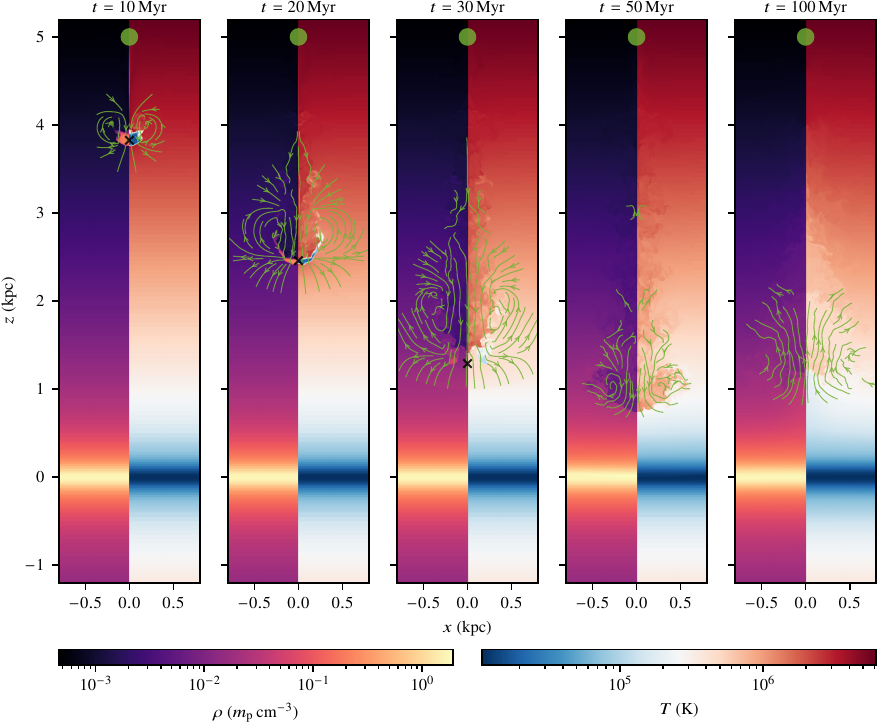}
    \caption{
    Snapshots of the interaction between the infalling cloud and the Galactic background medium from the adiabatic run, showing density (left half of each panel) and temperature (right half). 
    Each panel presents a cross-sectional slice perpendicular to the $y$-axis at the initial cloud centre. 
    Green streamlines visualize the flow pattern; a green disc marks the cloud's initial shape, size, and position, while a black cross indicates the bulk-cloud's instantaneous projected COM.
    The elapsed time since the start of the simulation is noted at the top of each panel.
    An animated version of this figure is available in the online supplementary material.
    }
    \label{im:timeseries_adiabatic}
\end{figure*}
The adiabatic run (opaque green curves in Fig.~\ref{im:comparison}) provides the baseline case in which the cloud evolves under hydrodynamics in a fixed gravitational potential, without cooling or conduction.

At early times ($t \lesssim \unit[10]{Myr}$), the cloud accelerates towards the Galactic mid-plane because its initial speed is below the local terminal speed at the starting height.
The motion is therefore approximately ballistic, and the COM trajectory and velocity remain close to the analytical and semi-analytical solutions (Fig.~\ref{im:comparison}a,b).
At the same time, the effective cloud radius increases (Fig.~\ref{im:comparison}c), reflecting Bernoulli-driven lateral expansion as the background flow is deflected around the moving cloud. 
Shear-driven KH instabilities develop from the outset due to the velocity shear at the cloud--halo interface. 
These initially appear as small ripples along the cloud surface, most prominently near the lateral flanks, and roll up into coherent vortical structures visible in the flow pattern (Fig.~\ref{im:timeseries_adiabatic}; green streamlines).
A narrow, dense tail forms directly behind the cloud, detaches after $\sim$\unit[10]{Myr}, and is subsequently dispersed over the next $\sim$\unit[20]{Myr} as the wake broadens and becomes increasingly turbulent.

The cloud reaches a maximum infall speed of $\sim$$\unit[135]{km\,s^{-1}}$ at $t\sim\unit[15]{Myr}$ (Fig.~\ref{im:comparison}b), after which ram-pressure drag becomes dynamically important and the cloud decelerates.
While KH instabilities continue to operate, the onset of deceleration triggers RT modes at the leading edge, where the effective acceleration opposes the direction of motion.
These grow most strongly along the symmetry axis and progressively disrupt the cloud from the front, producing finger-like structures between successive snapshots.
The morphology transitions into a head--tail configuration, with a fragmented leading region and a broad, vortical wake, giving rise to a jellyfish-like appearance in the density--temperature distribution (Fig.~\ref{im:timeseries_adiabatic}). 
The RT instability is fundamentally driven by the baroclinic generation of vorticity $\boldsymbol{\omega}$, where $\mathrm{D}\boldsymbol{\omega}/\mathrm{D}t \propto \nabla\rho \times \nabla p$. 
Given that the cloud's effective deceleration dynamically changes as it propagates through the stratified background medium in a spatially varying gravitational potential, a time-dependent treatment of the RT growth rate is necessary \citep[see][]{Schulreich:2022}.

A bow shock forms ahead of the cloud once the relative flow becomes transonic and briefly supersonic, as indicated by the COM Mach number evolution (Fig.~\ref{im:comparison}f).
The shock remains unsteady and evolves together with the cloud as the latter decelerates and fragments.
In this regime, analytical descriptions of the flow around a blunt body \citep[e.g.][]{Schulreich:2011} provide a useful framework for interpreting the structure of the shocked gas and the post-shock flow.

The most intense mass-loss phase coincides with the transition from accelerated infall to drag-dominated evolution. 
After an initial period of weak mass loss, the bulk-cloud mass begins to decrease rapidly once the instabilities reach non-linear amplitude, consistent with the onset time $t_\mathrm{M}$ of the mass-exchange model (Table~\ref{tab:mexparams}). 
This phase occurs when the cloud simultaneously attains a large effective cross section, experiences strong drag-induced deceleration, and is subject to fully developed KH and RT instabilities. 
Consistently, Fig.~\ref{im:comparison}h shows that the drag, convergence, and instability growth times become comparable during this stage, so that the cloud is disrupted on the same time-scale on which its global motion is being reshaped.

The velocity dispersions trace this evolution (Fig.~\ref{im:comparison}e): the diffuse component peaks first at $t\sim\unit[20]{Myr}$ with $\sigma_\mathrm{diff}\sim\unit[70]{km\,s^{-1}}$, when shear-driven stripping is strongest, followed by the mixing component at $t\sim 25$--$\unit[30]{Myr}$ with $\sigma_\mathrm{mix}\sim\unit[50]{km\,s^{-1}}$.
At approximately the same time, the bulk-cloud dispersion reaches its maximum of $\sigma_\mathrm{cl}\sim\unit[30]{km\,s^{-1}}$, indicating that the instabilities have grown to scales comparable to the cloud itself and are now disrupting the dense core.

The bulk-cloud mass effectively vanishes by $t\sim\unit[32]{Myr}$, when the density of the remaining material drops below the adopted bulk threshold (Fig.~\ref{im:comparison}d). 
This occurs at a COM height of $\sim$\unit[1.2]{kpc} (Fig.~\ref{im:comparison}a) and is consistent with the absence of the COM marker (black cross) in the last two panels of Fig.~\ref{im:timeseries_adiabatic}. 
During this transition, the diffuse component becomes dominant, while the mixing component traces gas that has been stripped from the cloud and subsequently diluted by interaction with the ambient medium. 
The mass in the mixing component continues to grow beyond the main disruption phase and reaches its maximum only at $t\sim\unit[55]{Myr}$, indicating that stripped material persists in an intermediate-density phase for an extended period before being fully mixed into the background. 
Over the full simulation time of $\unit[100]{Myr}$, the diffuse cloud mass increases to nearly 20 times the initial cloud mass (Fig.~\ref{im:comparison}d).

The effective radius reaches nearly twice its initial value around $t\sim\unit[20]{Myr}$ and subsequently declines as the cloud loses mass and fragments (Fig.~\ref{im:comparison}c). 
The inferred drag coefficient varies strongly throughout the evolution (Fig.~\ref{im:comparison}g), with a time-averaged value of $\overline{C_\mathrm{d}} \simeq 2.6 \pm 1.3$. 
This variability reflects the evolving effective shape of the cloud, from a compact, nearly spherical object to a flattened and highly irregular configuration. 
As long as a coherent bulk region exists, an effective drag coefficient can still be defined, albeit with large fluctuations, but it becomes ill-defined once the cloud is fully disrupted.

At late times, the cloud material is fully dispersed and forms a diffuse, turbulent plume. 
The bow shock, however, persists and propagates independently of the disrupted cloud, transitioning from a driven bow shock to a freely propagating disturbance with decreasing Mach number, and eventually crossing the Galactic mid-plane. 
As momentum is redistributed within the wake, the net downward motion progressively slows, stalls at $t\sim\unit[70]{Myr}$, and subsequently reverses. 
This behaviour is driven by the pressure deficit in the wake, which induces a recirculating backflow into the low-pressure region created by the cloud's passage. 
Although this motion does not represent a genuine fountain flow, it could observationally resemble a weak outflow component if inferred from kinematics alone.
The passage of the shock through the disc may further compress the ambient gas, potentially influencing local star formation, although such effects are beyond the scope of the present simulations.

Overall, the adiabatic run demonstrates that, in the absence of cooling, hydrodynamic instabilities, coupled to drag and deformation, rapidly destroy the cloud. 
The analytical and semi-analytical models provide useful reference solutions for the early evolution, while deviations become noticeable at later times. 
A more detailed comparison and discussion of the underlying causes is presented in Section~\ref{sec:comparison}.

\subsubsection{RC run}
\begin{figure*}
	\includegraphics[width=\textwidth]{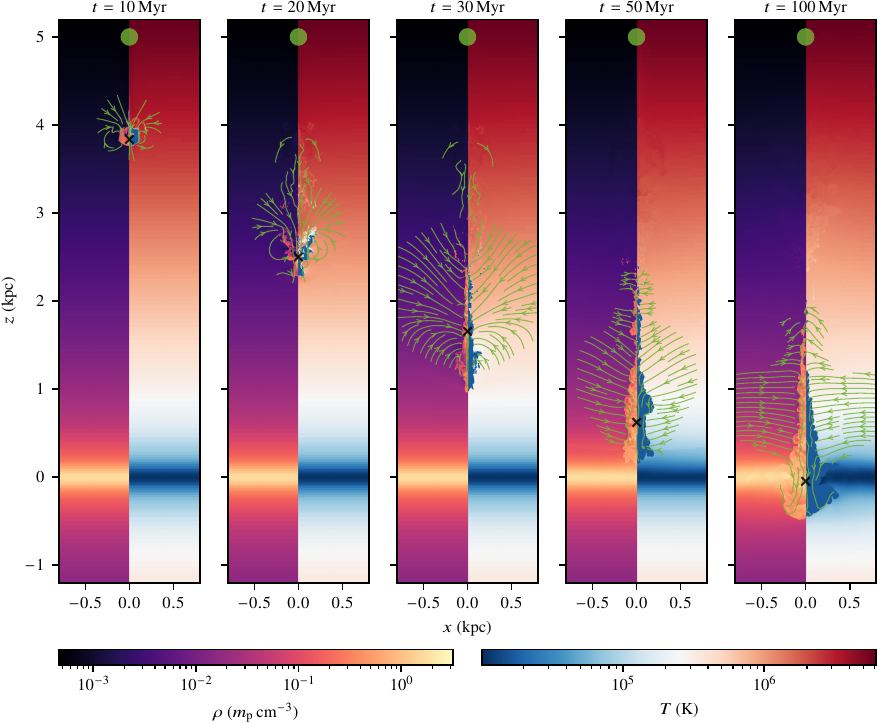}
    \caption{
    As for Fig.~\ref{im:timeseries_adiabatic}, but for the RC run.
    An animated version is available in the online supplementary material.
    }
    \label{im:timeseries_rc}
\end{figure*}
The inclusion of radiative cooling (opaque blue curves in Fig.~\ref{im:comparison}) leads to a qualitatively different evolution, allowing the cloud to survive the entire descent. For $t \lesssim \unit[10]{Myr}$, however, the behaviour remains close to the adiabatic case:
the cloud accelerates towards the Galactic mid-plane, with the COM trajectory and velocity following the analytical and semi-analytical solutions, while the effective radius increases due to Bernoulli-driven lateral expansion (Fig.~\ref{im:comparison}a--c).

The transition occurs once gas in the turbulent mixing layers cools efficiently.
Mixed material produced along the cloud--halo interface and within the wake undergoes thermal instability \citep{Field:1965} and rapidly cools towards the imposed temperature floor ($\sim$$\unit[10^4]{K}$), condensing into compact cloudlets and extended coherent structures that remain dynamically coupled to the flow.

A characteristic cloudlet size follows from the properties of these mixing layers as
\begin{equation}
\ell \sim \min(2 R_\mathrm{cl}, \sigma_\mathrm{mix} \tau_\mathrm{cool}) ,
\label{eq:ellrc}
\end{equation}
where $2 R_\mathrm{cl}$ sets the geometric limit imposed by the wake width, while $\sigma_\mathrm{mix} \tau_\mathrm{cool}$ measures the distance over which turbulent compression operates within one cooling time.
This yields typical sizes of a few parsecs, with a time-averaged value of $\sim$\unit[7]{pc}, indicating that the cloudlets are only marginally resolved at the highest grid level.

These dense structures are advected downstream, where they are channelled towards the symmetry axis and merge into a narrow filamentary tail (Fig.~\ref{im:timeseries_rc}).
The wake thereby organizes both mass and momentum, and the system develops into a two-component morphology consisting of a leading dense `parent' body and a continuously replenished filamentary tail.
Ongoing cooling in the mixing layers sustains this redistribution, while interactions between dense structures lead to repeated compressions and merging events, producing a highly structured flow.

This behaviour is reflected in the evolution of the effective radius (Fig.~\ref{im:comparison}c), which, after an initial growth phase, decreases to a local minimum corresponding to a transient, column-like configuration, and subsequently increases monotonically as a broad, dense head forms that is continuously supplied by cooled material from the wake.
By the end of the simulation, the effective radius exceeds its initial value by more than a factor of 3.5.

The mass evolution (Fig.~\ref{im:comparison}d) provides further quantitative support for this picture.
The bulk-cloud mass increases monotonically, indicating that condensation outweighs stripping.
The mixing component peaks at $t\sim\unit[25]{Myr}$ and subsequently declines as material is incorporated into the dense phase, while the diffuse component remains comparatively weak.

This condensation process enhances the dynamical coupling between the cloud and the ambient medium. 
As cooled gas accretes onto the cloud, it must be accelerated to the cloud velocity, resulting in an additional drag contribution associated with momentum loading (equation~\ref{eq:facc}). 
While ram-pressure drag reflects the momentum flux of the ambient flow past the cloud, this accretion-driven contribution arises from the direct incorporation of cooled material. 
As condensation proceeds, this contribution can become comparable to, or exceed, the classical ram-pressure drag, leading to a more efficient deceleration of the system.

As a consequence, the cloud remains close to a terminal-velocity-like regime over an extended period, following the local terminal-velocity solution over a substantial fraction of its trajectory (Fig.~\ref{im:comparison}b). 
In contrast to the adiabatic case, where rapid disruption prevents long-lived convergence, the sustained mass growth and enhanced drag allow the system to maintain this regime.

Hydrodynamic instabilities remain active but are fundamentally modified by cooling.
KH modes continue to drive mixing, while RT modes develop during deceleration; however, the cooled gas forms dense structures rather than dispersing.
The velocity dispersions in the diffuse and mixing components remain comparable to those in the adiabatic case, while the bulk-cloud dispersion reaches significantly higher values (Fig.~\ref{im:comparison}e), reflecting continuous compression, merging, and reorganization of dense gas.

Although the COM motion remains subsonic, the Mach number exhibits several distinct local maxima (Fig.~\ref{im:comparison}f).
The first enhancement at $t\sim\unit[20]{Myr}$ coincides with the formation of a bow shock ahead of the cloud, even though $\mathrm{Ma}<1$, indicating locally transonic conditions at the leading edge.
A second enhancement at $t\sim\unit[60]{Myr}$ occurs as the cloud approaches the cool disc gas, while a third at $t\sim\unit[85]{Myr}$ is associated with the interaction between the dense head and the trailing wake, which appears to push the head across the mid-plane. 
These features indicate that transonic conditions arise locally and intermittently, while the global flow remains subsonic.

The inferred drag coefficient varies strongly with time (Fig.~\ref{im:comparison}g), with a time-averaged value of $\overline{C_\mathrm{d}} \simeq 2.3 \pm 1.7$ over intervals in which the reconstruction remains physically meaningful. Temporary gaps occur when the reconstructed ram-pressure force becomes too small or slightly negative, preventing a reliable inversion. 
Since the deceleration parameter $\alpha$ is derived from $C_\mathrm{d}$, the corresponding terminal-velocity estimate exhibits analogous interruptions (Fig.~\ref{im:comparison}b).

The cooling time stays well below the free-fall time throughout (Fig.~\ref{im:comparison}h), keeping condensation active while the dynamical time-scales evolve as described above.
Overall, radiative cooling transforms the evolution from rapid disruption and mixing into a regime characterized by condensation, wake focusing, sustained mass growth, and enhanced dynamical coupling through accretion.

\subsubsection{RC+TC run}
\begin{figure*}
\includegraphics[width=\textwidth]{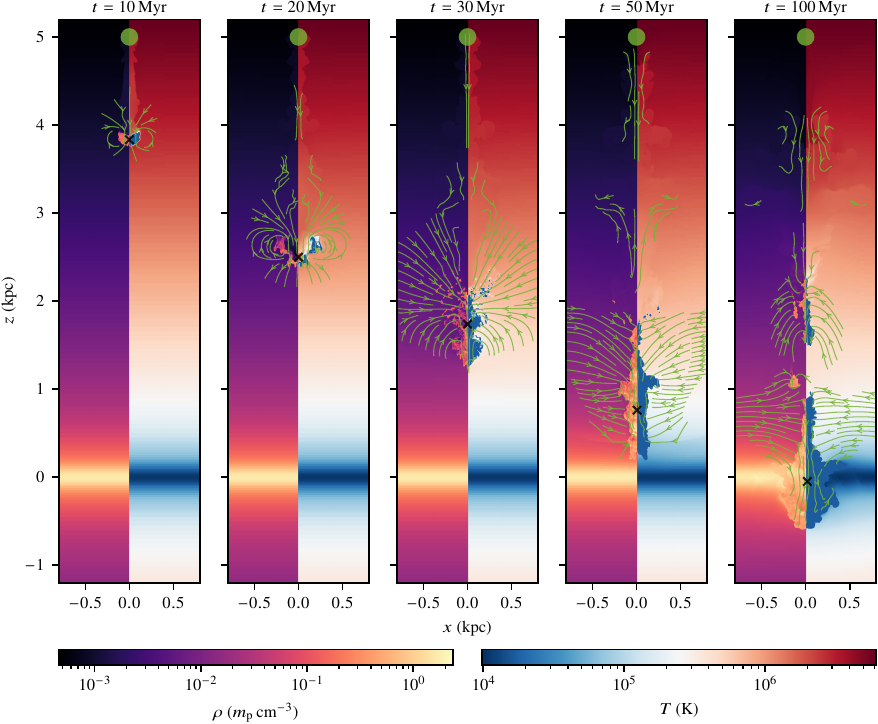}
    \caption{
    As for Fig.~\ref{im:timeseries_adiabatic}, but for the RC+TC run.
    An animated version is available in the online supplementary material.
    }
    \label{im:timeseries_rctc}
\end{figure*}
The inclusion of thermal conduction (opaque orange curves in Fig.~\ref{im:comparison}) leaves the global evolution broadly similar to the RC run. 
The cloud trajectory, velocity, Mach number (including transient compressions and weak shocks), and velocity dispersions closely follow those of the RC case throughout (Fig.~\ref{im:comparison}a,b,e,f). 
The bulk-cloud mass evolution is likewise nearly identical (Fig.~\ref{im:comparison}d), indicating that condensation remains efficient, while the diffuse and mixing components are slightly
enhanced. 
The inferred drag coefficient remains comparable, with a time-averaged value of $\overline{C_\mathrm{d}} \simeq 2.1 \pm 1.6$ (Fig.~\ref{im:comparison}g).

The primary differences arise in the structure of the turbulent mixing layers. 
Conductive heat transport smooths temperature and density gradients, broadening the interface between the cloud and the ambient medium and suppressing the growth of small-scale thermal instabilities.
As a result, the sharp boundaries and numerous compact cloudlets formed in the RC run are replaced by more coherent and spatially extended structures (Fig.~\ref{im:timeseries_rctc}).

This behaviour can be quantified by extending equation~\eqref{eq:ellrc} to include the Field length,
\begin{equation}
\lambda_\mathrm{F}=\sqrt{\frac{\kappa_\mathrm{eff,mix} T_\mathrm{mix}}{n^2_\mathrm{mix} \Lambda(T_\mathrm{mix},Z_\mathrm{mix})}} ,
\end{equation}
leading to
\begin{equation}
\ell \sim \max(\lambda_\mathrm{F}, \min(2 R_\mathrm{cl}, \sigma_\mathrm{mix} \tau_\mathrm{cool})) .
\end{equation}
In our simulations, $\lambda_\mathrm{F}$ spans a wide range, from $\sim$$\unit[2\times10^{-3}]{pc}$ up to a few parsecs, comparable to the numerical resolution limit. 
Thus, only the largest condensation scales are marginally resolved, while smaller scales remain unresolved. 
The resulting characteristic cloudlet sizes are $\ell\sim\unit[13]{pc}$ with substantial scatter.

The suppression of small-scale thermal instability manifests directly in the wake morphology. 
The filamentary tail becomes broader and less coherent, with gaps between individual condensations (Fig.~\ref{im:timeseries_rctc}), indicating reduced focusing efficiency and a larger fraction of intermediate-density material. 
The cloud is therefore organized into a set of dynamically related substructures rather than a single continuous filament, reminiscent of observed HVC complexes.

The effective radius evolution reflects this behaviour.
It follows the RC run at early times, including the initial Bernoulli-driven expansion, but exceeds it at later stages (Fig.~\ref{im:comparison}c), as reduced focusing leads to a broader spatial distribution of dense gas and hence a larger projected area.

The time-scale ratios (Fig.~\ref{im:comparison}h) show that the cooling time remains shorter than the other relevant time-scales, although it is slightly increased at early times compared to the RC run.
The evaporation time is initially comparable to the other time-scales, but subsequently increases strongly and exceeds all others by orders of magnitude, rendering it dynamically negligible over most of the evolution.

Overall, thermal conduction does not significantly alter the global dynamics in the present setup, but regulates the small-scale structure, leading to smoother interfaces, reduced fragmentation, and a broader, less coherent wake.

\subsubsection{Comparison with the analytical and semi-analytical model}
\label{sec:comparison}
\begin{figure*}
    \centering
    \includegraphics[width=0.94\textwidth]{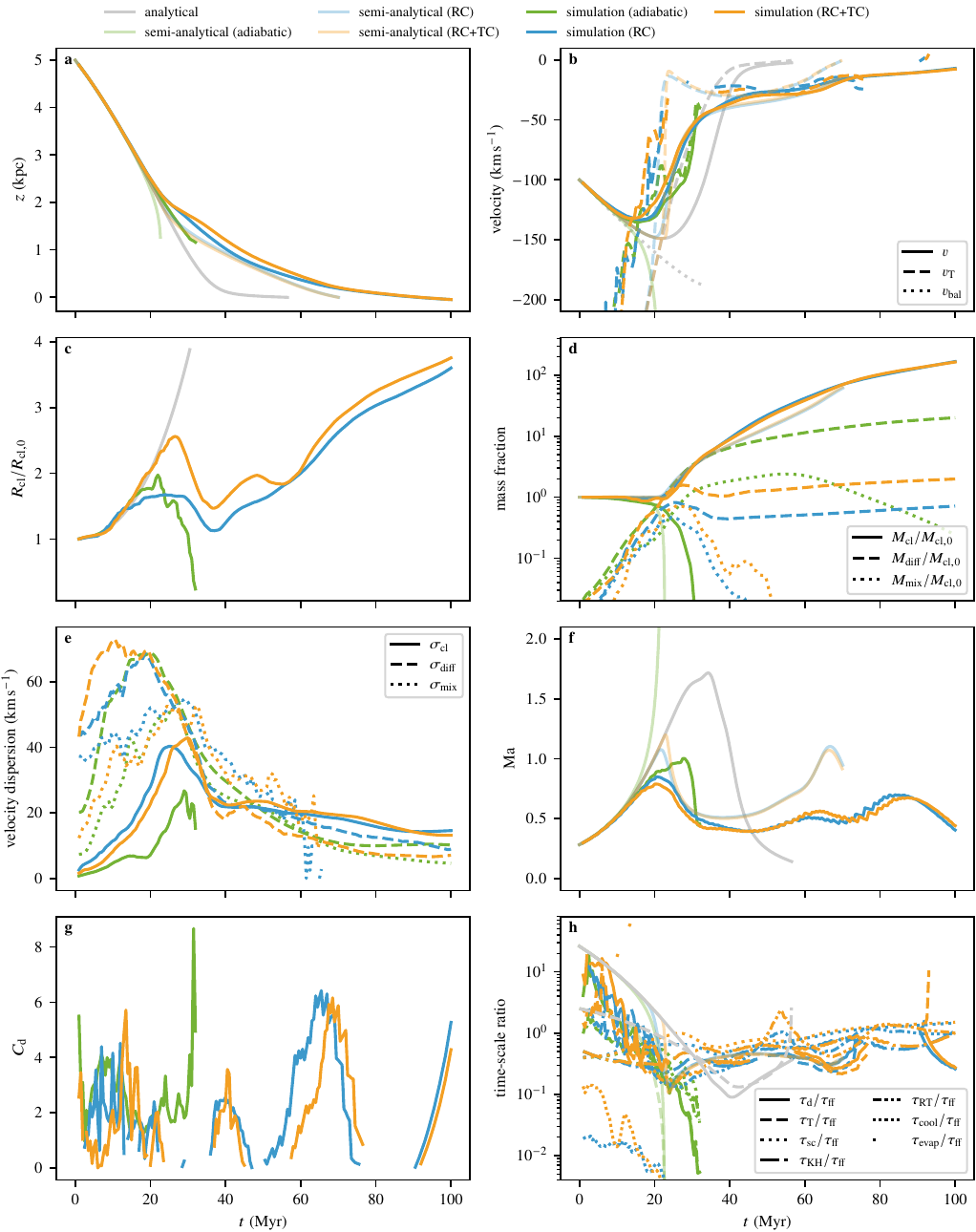}
    \caption{
    Time evolution of key dynamical and thermodynamic cloud properties.
    Coloured curves represent the simulations, semi-transparent coloured curves the semi-analytical solutions, and grey curves the analytical solutions (where available).
    The panels show:
    (\textbf{a})~COM vertical position;
    (\textbf{b})~COM vertical velocity, together with the corresponding terminal and ballistic solutions;
    (\textbf{c})~effective cloud radius, normalized to the initial radius;
    (\textbf{d})~mass of bulk, diffuse, and mixing components (normalized to the initial cloud mass);
    (\textbf{e})~velocity dispersion of the same components;
    (\textbf{f})~COM Mach number;
    (\textbf{g})~drag coefficient;
    (\textbf{h})~characteristic time-scales (drag, convergence, sound-crossing, KH/RT growth, RC, and evaporation), all normalized to the free-fall time.
    After the initial transient the dynamical ratios cluster around unity (strongly coupled evolution; see text), while $\tau_\mathrm{cool}$ lies below and $\tau_\mathrm{evap}$ rises well above the band.
    }
    \label{im:comparison}
\end{figure*}

\begin{table*}
\centering
\caption{Parameters of the mass-exchange model
(Section~\ref{sec:massex}) for the different simulation types.}
\label{tab:mexparams}
\begin{tabular}{lccccccc}
\hline
Simulation & $t_M$                  & $\Delta t$             & $\epsilon_{\rm strip}$ & $\epsilon_{\rm cond}$ & $\epsilon_{\rm evap}$ & $\eta$ & $q$ \\
           & ($\tau_\mathrm{KH,0}$) & ($\tau_\mathrm{KH,0}$) &                        &                       & ($10^{-6}$)           &        & \\
\hline
adiabatic  & 1.46                   & 0.38                   & 2.81                   & 0                     & 0                     &  --    & --   \\
RC         & 1.27                   & 0.08                   & 0.90                   & 2.08                  & 0                     & 16.74  & 5.51 \\
RC+TC      & 1.31                   & 0.02                   & 1.06                   & 2.29                  & 8.06                  & 6.54   & 4.18 \\
\hline
\end{tabular}
\end{table*}
The comparison between simulations and the (semi-)analytical solutions reveals a clear separation between regimes in which the underlying assumptions remain valid and those in which they break down (Fig.~\ref{im:comparison}). 
The analytical curves (grey) correspond to the constant-property limit, while the semi-analytical solutions additionally account for mass exchange. 
Both terminate once the cloud reaches the Galactic mid-plane.

A central result of the analytical parameter study (cf.~Section~\ref{sec:anaresults}) is that high-column-density clouds, such as the case considered here ($N_{\ion{H}{i}}=\unit[10^{20}]{cm^{-2}}$), approach the terminal velocity only very late along their trajectories, i.e.~shortly before disc crossing (Fig.~\ref{im:comparison}b). 
This reflects the large inertia per unit area of dense clouds and the correspondingly weak coupling to the ambient medium. 
The present setup therefore represents a particularly demanding case for terminal-velocity convergence.

In this context, the peak infall speed controls how rapidly the cloud traverses the halo and thus how much time is available to approach the
terminal-velocity solution. 
Compared to the corresponding constant-property solution, all simulations exhibit reduced peak infall speeds ($\sim$\unit[10]{per cent}; Fig.~\ref{im:comparison}b). 
This reduction arises from the breakdown of the constant-property assumption: in the simulations, the cloud undergoes mass loss, deformation, and an increase in effective cross section, which enhances the drag at earlier times and limits the maximum absolute velocity, even in the absence of cooling.

Radiative cooling modifies the subsequent evolution. In the RC and RC+TC runs, condensation allows the cloud to survive and introduces additional momentum loading, which prolongs the infall time and enables the system to remain close to the terminal-velocity solution over an extended period. 
This implies that lower-column-density clouds should reach terminal velocities even more readily.

The semi-analytical solutions reproduce the simulated evolution well in the cooling runs and initially also in the adiabatic case, but deviate systematically at later times in the latter. 
This behaviour is controlled by the mass-exchange parameters (Table~\ref{tab:mexparams}). 
In the cooling runs, condensation dominates ($\epsilon_{\rm cond}\gtrsim 2$) over stripping ($\epsilon_{\rm strip}\sim 1$), resulting in $\beta>0$ and an additional drag contribution through momentum loading. 
This maintains strong coupling between the cloud and the ambient medium and allows the semi-analytical trajectories to track the simulations closely. 
In contrast, the adiabatic case is characterized by strong stripping ($\epsilon_{\rm strip}\simeq 2.8$) and the absence of condensation ($\epsilon_{\rm cond}=0$ by construction), leading to $\beta<0$. 
For infall ($\varv<0$), the inertial term $-\beta \varv$ then acts in the direction of motion and effectively reduces the net drag, causing the semi-analytical solution to drift towards increasingly negative velocities (Fig.~\ref{im:comparison}b). 
The analytical solution, while not accounting for cloud disruption, does not exhibit this unphysical trend and therefore provides a closer approximation at late times.

Consistently, the onset time of mass exchange is similar in all cases, $t_M \sim \tau_{\rm KH,0}$, indicating that the transition to rapid mass evolution is set by the growth of KH instabilities. 
Thermal conduction contributes only weakly to the mass evolution, with $\epsilon_{\rm evap}\sim 10^{-5}$, implying that classical conductive evaporation is dynamically negligible.

The limitations of the analytical assumptions become apparent when considering the cloud geometry. 
While the Bernoulli solution initially follows the simulated radius evolution closely (Fig.~\ref{im:comparison}c), the simulated curves subsequently flatten and reach local maxima, whereas the analytical solution continues to increase. 
This reflects the breakdown of the quasi-steady and constant-property assumptions as the flow evolves.

Despite their simplifying assumptions, the semi-analytical solutions capture the global dynamics well in the cooling runs. 
In particular, both the drag coefficient and the cloud geometry are treated as constant in the model, whereas the simulations show that $C_\mathrm{d}$ is larger and strongly time-dependent (Fig.~\ref{im:comparison}g), and that the effective radius evolves significantly (Fig.~\ref{im:comparison}c). 
The good agreement therefore indicates that these effects enter only at higher order, while the dominant contribution to the dynamics arises from the mass evolution captured by the mass-exchange model.

Overall, the comparison shows that the predictive power of the semi-analytical model is fundamentally tied to its treatment of mass exchange. 
It provides an accurate description when condensation leads to net growth and sustained coupling, but becomes inadequate when mass loss reduces the effective drag and leads to progressively decoupled motion.

\section{Discussion}
\label{sec:discussion}

\subsection{Observable proxies}
To establish a direct link between the simulated gas dynamics and observable signatures, we derive a set of synthetic observables from the physically most complete RC+TC run.
Several such diagnostics were previously suggested as potential signatures of terminal-velocity-regulated halo-cloud infall by \citet{Benjamin:1999}, but have not yet been assessed in a fully dynamical cloud-evolution framework.

Many observational diagnostics are based on LOS integrated quantities. 
A fundamental example is the total hydrogen column density,
\begin{equation}
    N_{\mathrm H} = \int n_{\mathrm H} \dif s ,
\end{equation}
where $n_{\mathrm H}$ denotes the total hydrogen number density. 
It traces the projected gas distribution and provides a common basis for comparison with observations.

Depending on the diagnostic considered, different projection directions are adopted.

\subsubsection{VBs}
\label{sec:velocity_bridges}

In \ion{H}{i} observations of HVCs, VBs appear as emission features that connect gas over a wide range of radial velocities in position--velocity space \citep{Kalberla:2018}.
The term was first introduced by \cite{Verschuur:1969} in the context of interactions between IVCs and the Galactic disc. 
Subsequent surveys revealed numerous such structures linking HVCs and IVCs in \ion{H}{i} 21-cm emission \citep{Pietz:1996}.
Because they connect gas with substantially different radial velocities, VBs are commonly interpreted as signatures of dynamical interactions between clouds and their surrounding medium.
Observationally, the same cloud--ambient interaction that produces the VBs also leaves an SXR signature: regions of enhanced \emph{ROSAT} 1/4\,keV emission are found in spatial association with the HVC complexes \citep{Pietz:1996,Snowden:1995,Voges:1999}, which we interpret as compression of the ambient halo gas in Section~\ref{Sect:XRB}.
Other observational signatures of such interactions include head--tail morphologies.
A systematic census of northern-sky HVCs by \cite{Bruens:2000, Bruens:2001} showed that about \unit[20]{per cent} of the flux-limited sample of 252 HVCs exhibit such structures.
While head--tail features are generally interpreted as tracers of the motion of an HVC relative to the ambient medium, VBs extend over a substantially larger range in radial velocity and thus probe more extended kinematic interactions between the cloud and its environment.

To investigate whether analogous structures arise in our simulations, we construct synthetic position--velocity diagnostics from the cloud-related gas identified by the passive scalar field (region~$\mathcal C$; see Section~\ref{sec:regions}).
For this analysis we consider projections along the simulation $z$-axis, corresponding to a viewing geometry in which the cloud is observed along its direction of motion, i.e.~as if it were falling towards the observer.

\begin{figure}
    \includegraphics[width=\columnwidth]{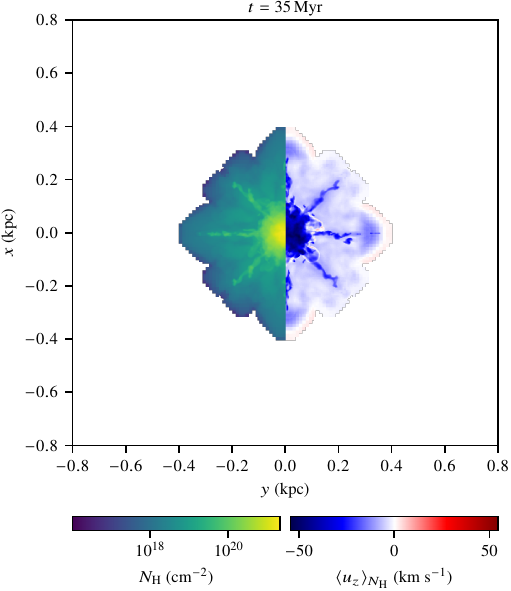}
    \caption{
    Total hydrogen column density (left half) and column-density-weighted LOS velocity (right half) of the cloud-related gas (region~$\mathcal{C}$; see Section~\ref{sec:regions}) in the RC+TC simulation at the indicated time, corresponding to a LOS along the $z$-axis.
    }
    \label{fig:velocity_bridge_xy}
\end{figure}

Figure~\ref{fig:velocity_bridge_xy} shows the projected distribution of the cloud-related gas after $\unit[35]{Myr}$ of evolution.
The left half of the figure presents the hydrogen column density, while the right half shows the corresponding column-density-weighted LOS velocity,
\begin{equation}
    \langle u_z \rangle_{N_{\mathrm H}} = \frac{1}{N_{\mathrm H}} \int n_{\mathrm H} u_z \dif z .
\end{equation}
The cloud appears as a compact high-column-density structure whose velocity field is dominated by coherent infall. 
Lower column-density material surrounding the cloud traces gas that has been stripped from the cloud and gas formed in situ within the wake, where mixing with the ambient medium leads to the formation of dense cloudlets that remain dynamically coupled to the flow.

As this material is transported downstream, it becomes increasingly organized as it converges towards the symmetry axis. 
This focusing is visible as narrow, radially oriented lanes of enhanced column density and elevated LOS velocity (Fig.~\ref{fig:velocity_bridge_xy}), which trace the combined inflow of stripped gas and condensed structures within the wake. 
The elevated velocities along these lanes indicate that they act as preferential inflow channels.

The highly regular, spoke-like appearance of these lanes suggests that this convergence is partially modulated by numerical effects. 
In particular, the Cartesian grid geometry favours preferential alignment along the grid axes and diagonals, while the fact that these features become markedly more pronounced only when thermal conduction is included suggests that the operator-split treatment of conduction may contribute to the enhanced anisotropy seen in the RC+TC run.
We therefore interpret the lane-like structures as a superposition of genuine physical wake focusing and numerically induced modulation. 
While their detailed morphology is likely affected by these numerical effects, the structures nevertheless trace the large-scale convergence of cloud material within the wake.

This spatial organization of the wake directly imprints on the velocity structure of the gas. 
For each projected position $y$, the gas density is integrated along the LOS ($z$) and across the transverse $x$-direction, while being binned by its LOS velocity.
This yields a discretized position--velocity intensity distribution corresponding to the bin-averaged form of
\begin{equation}
I(y,u_z) = \iint n_{\mathrm H} \delta\!\left(u_z - u'_z\right) \dif x \dif z ,
\label{eq:pv_intensity}
\end{equation}
where $\delta$ denotes the Dirac delta function and the averaging is performed over finite velocity bins in $u_z$.
If thermal Doppler broadening is included, the delta function is replaced by a Gaussian line profile
\begin{equation}
    \phi(u_z-u'_z) = \frac{1}{\sqrt{2\uppi} \sigma_{\mathrm{th}}} \exp\!\left[-\frac{(u_z - u'_z)^2}{2 \sigma_{\mathrm{th}}^2}\right] ,
    \label{eq:doppler}
\end{equation}
with thermal velocity dispersion $\sigma_{\mathrm{th}} = \sqrt{k_{\mathrm B} T / m_{\mathrm p}}$.

\begin{figure*}
    \includegraphics[width=\columnwidth]{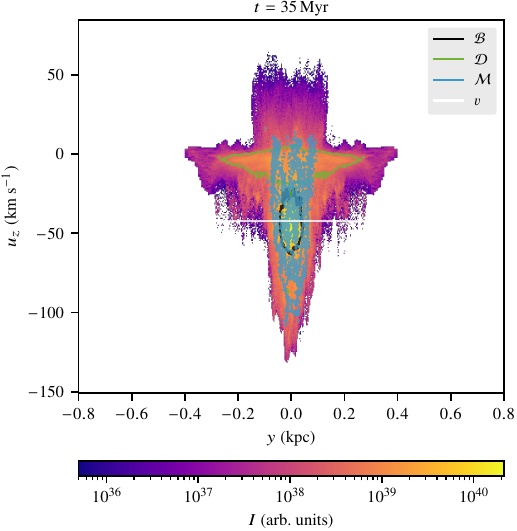}
    \includegraphics[width=\columnwidth]{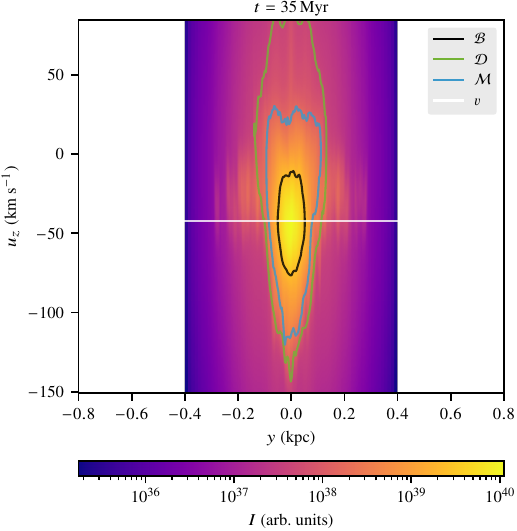}
    \caption{
    Synthetic position--velocity diagrams of the cloud-related gas (region~$\mathcal{C}$; see Section~\ref{sec:regions}) in the RC+TC simulation at the indicated time.
    The left panel shows the ideal case without thermal broadening, whereas the right panel includes thermal Doppler broadening.
    The colour scales represent the intensity distributions defined by equations~\eqref{eq:pv_intensity} and \eqref{eq:doppler}.
    Contours trace the contributions from the bulk cloud (black), diffuse cloud-related gas (green), and the mixing region (blue), each plotted at \unit[20]{per cent} of its respective maximum intensity.
    The horizontal white line indicates the density-weighted bulk-cloud velocity.
    }
    \label{fig:velocity_bridge_pv}
\end{figure*}

The resulting position--velocity diagrams are shown in Fig.~\ref{fig:velocity_bridge_pv}. 
In both panels the emission forms a continuous structure that connects gas near the bulk-cloud velocity with gas at substantially different velocities, producing a clear VB.

The structure is not uniform but exhibits a compact high-intensity core associated with the bulk cloud, surrounded by more extended emission at lower intensities. 
The latter traces gas that has been stripped from the cloud or formed within the wake and subsequently accelerated, decelerated, or mixed within the flow, yielding a continuous distribution in velocity space. 
The overplotted contours confirm that the bridge is primarily composed of diffuse and mixing gas, while the bulk component remains confined to a narrow velocity range.

Comparing the two panels shows that thermal broadening mainly smooths the velocity distribution and reduces small-scale structure, but does not alter the overall morphology of the bridge, indicating that the feature reflects the underlying kinematics rather than being an artefact of limited velocity resolution. Such structures are consistent with the VBs observed in \ion{H}{i} data.

Beyond VBs and head--tail morphologies, the internal density structure of the cloud provides a third, complementary constraint on its dynamical state. 
As noted in Section~\ref{sec:bernoulli}, ram-pressure compression of the windward face builds up a steep, quasi-exponential density gradient along the direction of motion, whose scale length is set by the balance between the external dynamic pressure and the internal pressure response. 
To leading order, hydrostatic balance in the cloud frame implies a compression scale length $L_\mathrm{c}\sim c_\mathrm{s}^{2}/|\dot\varv|$, so that $L_\mathrm{c}$ shortens as the instantaneous deceleration $|\dot\varv|$ increases; the integrated column-density asymmetry between the leading and trailing faces correspondingly encodes the cumulative momentum exchange. 
In our simulations this axial compression is clearly present at the leading edge during the drag-dominated phase, although in the cooling runs it is partly masked by the condensation of cooled material into the head and wake. 
In principle, spatially resolved \ion{H}{i} or absorption mapping of the density profile across an HVC head -- particularly for clouds with a well-defined head--tail axis -- could therefore be used to estimate the instantaneous (from the local scale length) or time-integrated (from the head--tail column-density contrast) deceleration, offering an independent, geometry-based probe of the drag history that complements the purely kinematic VB and head--tail diagnostics.

\subsubsection{DGR and optical extinction}
\label{sec:extinction}
Another observable consequence of the interaction between halo clouds and the ambient medium is the modification of the dust content and the resulting optical extinction.
As the cloud mixes with the surrounding gas, the DGR of the cloud material is expected to change and may therefore provide an observational tracer of the interaction.

In the Milky Way disc the DGR is well constrained, with a characteristic value of $\mathrm{DGR}_{\mathrm{ISM}} \simeq 1/100$ by mass.
In the Galactic halo the dust abundance is expected to be significantly lower due to efficient dust destruction by thermal sputtering, grain--grain collisions, UV radiation, cosmic rays, and shocks driven by stellar feedback.
Since the exact halo value remains poorly constrained, we adopt the disc value $\mathrm{DGR}_{\mathrm{ISM}}$ throughout the computational domain, which provides an upper limit for the dust content and thus for the resulting optical extinction.

To estimate the DGR associated with the cloud we consider gas within region~$\mathcal{V}(t)$, defined by the local flow-speed criterion $|\mathbfit{u}|\ge\unit[5]{km\,s^{-1}}$, thereby excluding the quiescent background medium.
This definition differs from the tracer-based classification introduced in Section~\ref{sec:regions} and ensures that dynamical structures such as bow shocks are included even if they do not contain significant amounts of tracer material.

We assume that the cloud is initially dust-free, corresponding to an idealized halo cloud of external or weakly processed origin.
Under this assumption, dust is contributed exclusively by ambient gas mixed into the cloud, so that the local dust content is proportional to the ambient gas fraction, given by $(1-C)$.
The column-integrated DGR of the material in region~$\mathcal{V}(t)$ then amounts to
\begin{equation}
    \mathrm{DGR}_{\mathcal{V}} = \frac{\Sigma_\mathcal{V}^{\mathrm{dust}}}{\Sigma_\mathcal{V}^{\mathrm{gas}}} 
    = \frac{\int_{\mathcal{V}(t)} (1-C) \mathrm{DGR}_{\mathrm{ISM}} \rho \dif s}{\int_{\mathcal{V}(t)} \rho \dif s} ,
\end{equation}
where $\Sigma_\mathcal{V}^{\mathrm{dust}}$ and $\Sigma_\mathcal{V}^{\mathrm{gas}}$ denote the dust and gas column densities, respectively.
From this, we estimate the corresponding optical extinction as
\begin{equation}
    A_V = \frac{\mathrm{DGR}_{\mathcal{V}}}{\mathrm{DGR}_{\mathrm{ISM}}}
    \left(
    \frac{N_{\mathrm H}}{\unit[2.21 \times 10^{21}]{cm^{-2}}}
    \right)\,\mathrm{mag} ,
\end{equation}
which adopts a scaling motivated by the empirical relation of \citet{Guever:2009}, recovered exactly when the DGR of the material in region~$\mathcal{V}$ equals that of the ISM.

A key assumption of this approach is that dust dynamics follows the gas dynamics, as represented by the passive scalar field.
This implies that dust grains are well coupled to the gas, i.e.~that their stopping time due to drag forces is much shorter than the characteristic flow time-scale.
This approximation is valid for small grains (sub-micron sizes, typically $\lesssim$\unit[0.1]{\micron}), which are effectively entrained in the gas flow, whereas larger grains may partially decouple.

\begin{figure}
    \centering
    \includegraphics{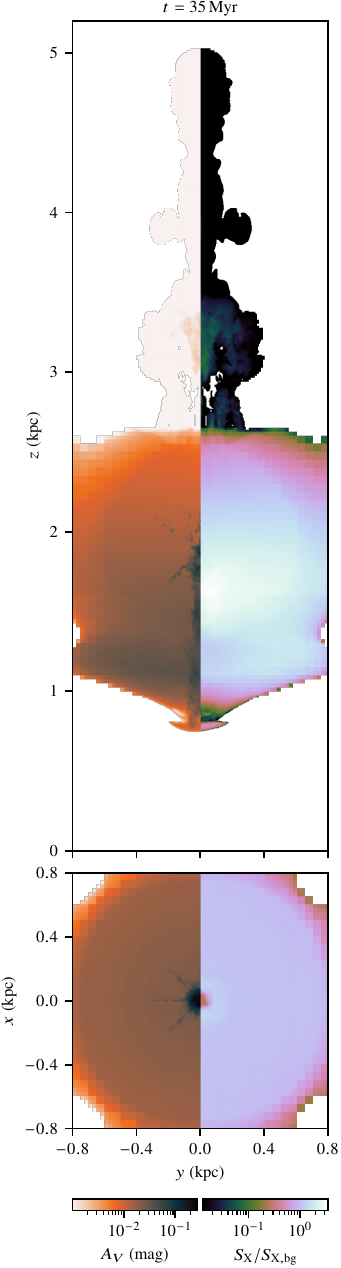}
    \caption{
    Optical extinction (left half) and SXR surface brightness in the \emph{ROSAT} C-band (right half; shown relative to its median background level) of the dynamically disturbed gas (region~$\mathcal{V}$; see Section~\ref{sec:extinction}) in the RC+TC simulation at the indicated time.
    The top panel corresponds to a LOS along the $x$-axis, while the bottom panel shows the projection along the $z$-axis.
    }
    \label{im:synthobs}
\end{figure}

To illustrate the expected extinction signal, we generate synthetic extinction maps (left half of Fig.~\ref{im:synthobs}) from the same simulation snapshot used for the VB analysis in Section~\ref{sec:velocity_bridges}.
Two viewing geometries are considered.
The first, shown in the bottom panel, corresponds to the vertical projection introduced before,
where the cloud is observed along its direction of motion.
The second geometry considers the same cloud located at the same Galactocentric radius but on the opposite side of the Galaxy, corresponding to a distance $d=2 R_0 \simeq \unit[16.2]{kpc}$ from the Sun along the Galactic plane (top panel of Fig.~\ref{im:synthobs}).

For this second configuration the column densities are computed by integrating along the $x$-direction rather than along the true LOS from the Sun.
This introduces a small geometric approximation because the actual LOS towards the cloud would be inclined relative to the Galactic plane.
Given the maximum cloud height of $z_0=\unit[5]{kpc}$, the resulting deviation corresponds to a relative path-length error of at most $1 -d / \sqrt{d^2 + z_0^2} \simeq \unit[4.4]{\text{per cent}}$.
However, given the much larger uncertainties in the dust content and mixing efficiency, this geometric effect represents only a minor contribution to the overall uncertainty in the computed extinction.

Figure~\ref{im:synthobs} shows that the extinction signal is highly structured and closely traces the distribution of dynamically disturbed gas. 
In both projections the strongest extinction coincides with the dense, compressed head of the cloud and with the trailing wake. 
This does not reflect dust intrinsic to the cloud -- which is dust-free by assumption -- but dust-bearing ambient material that has been mixed in and, in the cooling run, condensed onto the cloud, and is then concentrated at high column density around the compressed head and along the wake. 
The diffuse, pristine cloud gas, by contrast, remains characterized by very low extinction.

The morphology of the extinction maps reflects the underlying flow structure. 
In the vertical projection, the wake appears as an extended and centrally concentrated structure due to projection along the direction of motion, whereas in the edge-on view it forms a broad, vertically extended distribution trailing behind the cloud. 
In both cases, the extinction is largest where mixing and condensation are most
efficient, indicating that the dust content traces the dynamical interaction rather than the original cloud material.

Overall, the extinction signal remains modest, with typical values of $A_V \lesssim \unit[10^{-1}]{mag}$ (Fig.~\ref{im:synthobs}), implying that such clouds would be difficult to detect in optical extinction alone. 
Nevertheless, the spatial distribution of $A_V$ provides a sensitive tracer of mixing and mass exchange in the wake.

This mixing-driven scenario offers a natural interpretation of an otherwise puzzling observational result. 
\citet{Hernandez:2013} found that the intermediate-velocity molecular cloud IV21 (IVC\,135+54$-$45), located $\sim$\unit[300]{pc} above the disc, has a distinctly sub-solar metallicity ($\log(Z/Z_{\sun})=-0.43\pm0.12$) -- the first such cloud identified in the solar neighbourhood -- yet is bright in far-infrared emission and rich in molecular (CO) gas, an atypical combination for low-metallicity gas; its $\unit[100]{\micron}$ intensity per unit \ion{H}{i} column, $I_{\unit[100]{\micron}}/N_{\ion{H}{i}}$, lies a factor of $\sim$3 below the local high-latitude mean. 
They interpreted IV21 as the signature of an infalling, low-metallicity cloud mixing with disc gas. Our simulations make this interpretation quantitative: a dust-poor (here dust-free) infalling cloud acquires dust precisely where it mixes with the comparatively dust-rich ambient medium, producing a spatially variable DGR in which low-metallicity gas and locally enhanced dust coexist, exactly the combination observed in IV21. 
The model thus predicts that the dust content of such clouds should correlate spatially with kinematic tracers of mixing (the VBs and disturbed gas of Section~\ref{sec:velocity_bridges}) rather than with the pristine cloud body.

These predictions are becoming testable through the rapid progress in 3D dust mapping of the solar neighbourhood \citep[e.g.][]{Edenhofer:2024}. 
In particular, \citet{ONeill:2026} have carried out the first 3D spatial search for anomalous-velocity clouds at the local disc--halo interface, combining a parsec-resolution dust map with \ion{H}{i} kinematics to recover distances, 3D morphologies, and DGRs for a sample of high-altitude clouds, and find that classical IVCs constitute only $\sim$18 per cent of these, the remainder being low-radial-velocity structures. 
A detailed, quantitative confrontation between these data and our models -- both analytical and hydrodynamical -- is beyond the scope of the present paper and is best pursued separately. 
We note, however, that several of the high-altitude clouds they reconstruct in 3D (such as Draco and IVC\,135) display elongated, segmented structures that are qualitatively reminiscent of the head--tail and filamentary wakes produced in our cooling runs (Figs~\ref{im:timeseries_rc} and \ref{im:timeseries_rctc}), although not generally aligned with the normal to the Galactic plane, as expected once orbital and rotational motions tilt the infall direction away from the vertical. 
Notably, the cloud IVC\,135 is the same object as IV21 above -- the \citet{Hernandez:2013} cloud -- so the metallicity--dust puzzle and the new 3D morphology refer to a single, well-characterized cloud, an attractive test case for the mixing scenario. 
A dedicated comparison of the predicted internal density structure, DGR variations, and velocity bridges with these 3D reconstructions is a promising direction for future work.

\subsubsection{SXR surface brightness}
\label{Sect:XRB}

SXR emission provides a complementary probe of the interaction between halo clouds and the ambient medium.
As the cloud moves through the halo, it compresses and perturbs the surrounding gas, leading to enhanced emission measures that can be detected in SXR bands.

Observational evidence for such excess emission has been reported from the \emph{ROSAT} all-sky survey, which provides sensitivity to SXRs below the carbon K-shell ($E\simeq \unit[0.284]{keV}$; hereafter C-band). 
Several studies found that parts of northern HVC complexes are associated with enhanced C-band emission \citep{Kerp:1994,Herbstmeier:1995,Kerp:1999}.
These X-ray excesses are not exactly coincident with the \ion{H}{i} column-density maxima but are spatially offset, and not every \ion{H}{i} maximum shows an excess, likely reflecting the line-of-sight depth structure of the clouds.
While early interpretations attributed this excess to additional heating mechanisms such as magnetic reconnection, a combined analysis of the C-band and higher-energy M-band emission demonstrated that the observed signal is more naturally explained by an increased emission measure due to compression of the ambient coronal gas \citep{Kerp:1998}.
The absence of a corresponding excess in the M-band indicates that the plasma temperature remains largely unchanged, supporting a compression-dominated origin of the emission.

A further argument confirms that the excess reflects genuine emission rather than an absorption artifact. In the \emph{ROSAT} C-band, the foreground photoelectric absorption is dominated by helium rather than by the more abundant hydrogen \citep{Wilms:2000}. Because the helium and hydrogen abundances are fixed by primordial nucleosynthesis -- and helium, as a noble gas, undergoes no chemistry that would alter its abundance in the ISM -- the excess cannot be mimicked by a local deficit in their abundances reducing the foreground absorption.

To assess the expected X-ray signature of the dynamically disturbed gas (region~$\mathcal{V}$; see Section~\ref{sec:extinction}), we compute synthetic surface-brightness maps in the \emph{ROSAT} C-band (0.11--$\unit[0.28]{keV}$) from the same simulation snapshot used in the previous sections.
Assuming optically thin plasma emission, the surface brightness is obtained by integrating the band-limited emissivity $\varepsilon_\mathrm{X}$ along the LOS,
\begin{equation}
    S_\mathrm{X} = \frac{1}{4 \uppi} \int \varepsilon_\mathrm{X}(\rho,T,Z) \dif s .
\end{equation}

The right half of Fig.~\ref{im:synthobs} shows the resulting C-band surface brightness for the two viewing geometries introduced above.
For visualization, the surface brightness is shown relative to a background level defined by the median of the distribution within region~$\mathcal{V}$, thereby highlighting relative enhancements due to dynamical compression.

In the edge-on projection (top panel), enhanced emission is clearly visible ahead of the cloud, where the ambient gas is compressed by the cloud's motion. 
This emission forms an extended, bow-shaped structure that traces the interaction between the cloud and the surrounding medium. 
The brightness distribution exhibits a characteristic high--low--high pattern along the direction of motion: 
a bright compression region at the leading edge, followed by a zone of reduced emission associated with the expansion and relaxation of the compressed halo gas downstream of the bow-shock region, and a subsequent increase further downstream due to the integrated contribution of hot halo gas along the LOS.

The downstream region is therefore more challenging to interpret.
Although substantial emission is present, it largely reflects the LOS integration of hot halo gas rather than a distinct signature of the wake itself. 
As a result, the structured distribution of cooled and condensed material within the wake, which is evident in the density field, is only weakly reflected in the SXR maps.

In contrast, the vertical projection (bottom panel) integrates along the direction of motion and therefore does not preserve a clear separation between upstream and downstream regions. 
Instead, the emission appears more symmetric and centrally concentrated, reflecting
the superposition of compressed gas in front of the cloud and more diffuse material in its wake.

In both cases, the emission is strongest in regions of enhanced density rather than elevated temperature, indicating a compression-dominated origin. 
Accordingly, no corresponding enhancement is found in the M-band (0.44--\unit[1.21]{keV}).

This behaviour agrees with the observational interpretation of \citet{Kerp:1998} and supports a scenario in which SXR emission arises primarily from compression of the ambient medium rather than from additional heating mechanisms.


\subsection{Limitations and future work}
\label{sec:limitations}

The present simulations rely on a number of simplifying assumptions regarding both the large-scale Galactic environment and the internal structure of the cloud.

On large scales, we neglect the effects of galactic rotation, which introduces both radial shear and vertical lag in extraplanar gas \citep[e.g.][]{Kalberla:2009}. 
Using the Oort constants derived from \emph{Gaia}~DR2 \citep{Gaia:2018}, $A=~\unit[15.1]{km\,s^{-1}\,kpc^{-1}}$ and $B=\unit[-13.4]{km\,s^{-1}\,kpc^{-1}}$ \citep{Li:2019}, the local radial velocity gradient at the solar circle is $\partial V_\mathrm{circ}/\partial R|_{R_0}=-(A+B)=-\unit[1.7]{km\,s^{-1}\,kpc^{-1}}$. 
This corresponds to a velocity difference of $\unit[2.7]{km\,s^{-1}}$ across the $\unit[1.6]{kpc}$ wide computational domain and a displacement of $\unit[278]{pc}$ over the simulation time of $\unit[100]{Myr}$. 
Since the cloud is much smaller than the computational domain, shear across the cloud remains minimal.
Vertical shear can be larger at high altitudes, with an observed velocity gradient of $\unit[-15]{km\,s^{-1}\,kpc^{-1}}$ \citep{Marasco:2011}, corresponding to a velocity difference of up to $\unit[75]{km\,s^{-1}}$ at the initial height $z_0=\unit[5]{kpc}$. 
However, this difference continuously decreases as the cloud approaches the disc, and the internal velocity dispersion of both the cloud and its wake, of order tens of km\,s$^{-1}$, ensures that any residual shear is efficiently washed out by turbulence.
Furthermore, we take the background medium to be static ($\varv_\mathrm{bg}=0$) and initially unperturbed, so that the cloud falls into fresh, undisturbed halo gas throughout its trajectory. We thereby neglect any pre-existing structure in the ambient medium -- for example gas already stirred or enriched by earlier infalling clouds or by galactic-fountain activity -- that a continuously processed halo would contain.
While these assumptions provide a controlled environment for isolating the cloud--halo interaction, they neglect large-scale shear flows and global orbital evolution that may influence the dynamics over longer time-scales.

A related idealization concerns the thermal treatment of the background. 
As discussed in Section~\ref{sec:domain}, near $z\sim\unit[1]{kpc}$ the hot halo would, in isolation, be thermally unstable, while in a real galaxy it is held close to a statistical equilibrium by feedback processes -- heating, galactic-fountain cycling, and turbulence -- that we do not model. 
We mimic this maintained state by restricting radiative cooling and conduction to cloud-tagged gas (Section~\ref{sec:physics}), which isolates the cloud-driven mixing and condensation from spurious, volume-filling cooling of the halo but also means that the gas condensing onto the cloud in the cooling runs is, by construction, gas that would not have cooled spontaneously in the absence of the cloud. 
A genuinely turbulent, multiphase halo could develop its own cold structures and pressure fluctuations, modifying both the supply of condensable material and the effective density contrast experienced by the cloud, and hence the calibration of the condensation efficiency $\epsilon_{\mathrm{cond}}$ (and, to a lesser extent, $\epsilon_{\mathrm{strip}}$) in our mass-exchange model. 
We therefore regard our parametrization as appropriate for clouds moving through a quasi-stationary, statistically maintained hot halo; its extension to a self-consistently multiphase, thermally unstable background -- where mass exchange may depend on the ambient state as well as on the cloud properties -- is an important avenue for future work. 
Another, more pragmatic reason for treating the background as simply as possible here is that it would be virtually impossible, in a first thorough analysis of the cloud dynamics and its internal structure, to disentangle cloud and background effects cleanly.

A further idealization is our restriction to purely vertical infall at fixed Galactocentric radius $R$.
Because the Galactic potential is axisymmetric and deepens towards the centre, it is not separable in $r$ and $z$ -- its cross derivative $\partial^2\Phi/(\partial r\partial z)\neq 0$ couples the radial and vertical motion -- so that a real cloud would acquire a growing radial velocity component and drift inward as it falls.
This drift has two consequences.
Dynamically, it strengthens the drag: both the background density and the gravitational acceleration rise towards the Galactic centre, and the resulting increase in ram pressure makes disruption before disc-crossing more likely.
Kinematically, the asymmetric pressure distribution set up by the Bernoulli effect and ram pressure exerts a torque on the cloud, while baroclinic vorticity generation from the misalignment of density and pressure gradients ($\nabla\rho\times\nabla p\neq 0$) at the cloud--halo interface further drives internal rotation and turbulent fragmentation. A detailed treatment of these effects is deferred to a future paper.

The initial cloud is assumed to be homogeneous and approximately spherical, which facilitates direct comparison with the analytical and semi-analytical models. 
Real HVCs, however, are expected to exhibit internal density structure and multiphase substructure, which can affect their stability, mixing behaviour, and mass exchange with the ambient medium \citep[e.g.][]{Sander:2021}. 
In particular, pre-existing inhomogeneities may seed instabilities and modify the efficiency of cloudlet formation and wake focusing.

On smaller scales, we do not include magnetic fields. 
In a magnetized medium, magnetic draping around the cloud can lead to the formation of a compressed field layer at the leading edge, which may stabilize the interface and suppress the growth of KH and RT instabilities.
This can reduce mixing and alter the morphology of the wake, while magnetic pressure and tension provide an additional dynamical contribution. 
Recent magnetohydrodynamical studies further indicate that magnetic draping can introduce an additional drag component and enhance mass growth by increasing the coupling between the cloud and the ambient medium \citep{Kaul:2025}. 
Magnetic fields also render thermal conduction anisotropic, with heat transport proceeding preferentially along field lines. 
While our simulations allow for saturated thermal conduction, the absence of magnetic fields implies that the effective conductivity perpendicular to the field may be overestimated. 
To partially account for this effect, we employ a reduction factor for the conductivity; however, this approach can only provide an approximate description of a fully magnetized plasma.
Previous studies have shown that saturated thermal conduction can significantly alter the evolution of HVCs by promoting condensation rather than evaporation and by stabilizing the cloud against disruption, thereby modifying the efficiency of mixing and mass exchange \citep[e.g.][]{Sander:2023}. 
In addition, we neglect explicit viscosity.
The role of viscosity in diffuse astrophysical plasmas remains uncertain, as the effective viscosity depends on poorly constrained microphysical processes such as small-scale turbulence and particle interactions.
While viscosity can influence the development of instabilities at cloud interfaces \citep[e.g.][]{Roediger:2013}, previous studies indicate that turbulent mixing dominates over viscous dissipation in the low-density halo environment considered here.

Self-gravity and star formation are not included. The initial cloud mass is well below its theoretical Jeans mass ($M_\mathrm{J} \sim \unit[5 \times 10^8]{M_{\sun}}$), such that large-scale gravitational collapse is not expected. 
Even in simulations that include radiative cooling, where the cloud mass can grow by up to two orders of magnitude, it remains at most $\sim$$\unit[10^6]{M_{\sun}}$, and thus well below the threshold for significant gravitational collapse. 
However, self-gravity may still influence the evolution of dense structures within the cloud and its wake, potentially stabilizing cloudlets and enhancing their accretion
efficiency. 
Additionally, gravitational focusing may augment the mass flux channelled into the cloud's turbulent wake.
Recent observations provide direct evidence that star formation can occur within HVCs under suitable conditions, likely triggered by cloud--cloud interactions and gas compression \citep{He:2026}. 
Furthermore, the simulations assume optically thin conditions and neglect radiative self-shielding.
Self-shielding matters in dense, optically thick gas, where it suppresses the local ionizing flux and helps maintain a cold, neutral phase. 
In our setup, this cold neutral phase is produced directly by the radiative cooling already included (with an imposed temperature floor). 
Although the ambient ionizing field strengthens as the cloud descends towards the disc, the cloud simultaneously becomes denser and more optically thick, so that in reality it would self-shield efficiently in precisely those regions and remain largely neutral, consistent with our neutral-gas treatment. 
Neglecting explicit self-shielding is therefore unlikely to affect the bulk dynamical and thermal evolution, although it may alter the detailed ionization structure of the lower-density envelope and wake, particularly near the disc where the radiation field is strongest.

A further limitation concerns the sampling of the cloud parameter space. 
The numerical experiments presented here adopt a single, high initial column density ($N_{\ion{H}{i}}=\unit[10^{20}]{cm^{-2}}$; Section~\ref{sec:numresults}), and the mass-exchange parameters of Table~\ref{tab:mexparams} are calibrated for this case alone. 
There is good reason to expect these parameters to depend systematically on column density. 
Lower-column-density clouds have a larger surface-to-mass ratio and a smaller density contrast, so that KH stripping removes a larger mass fraction per dynamical time and the cooling time of the mixed gas changes relative to the (shorter) disruption time; both effects shift the balance between the condensation and stripping terms ($\epsilon_{\mathrm{cond}}$, $\epsilon_{\mathrm{strip}}$) and may move a cloud across the critical-ablation threshold of \citet{Marinacci:2010} that separates net growth from net loss. 
The present run should therefore be viewed as the first of a planned series of `falling-cloud' experiments -- a stratified-background counterpart to the cloud-in-wind tunnel studies -- designed to empirically characterize the effective drag coefficient and the mass-exchange parametrization as functions of cloud column density, size, metallicity, and infall velocity. 
Mapping out this dependence is a natural next step towards a predictive, observationally calibrated description of halo-cloud evolution.

\section{Conclusions}
\label{sec:conclusions}

In this work, we investigated the dynamics of IVCs and HVCs moving vertically through a stratified Galactic background medium and reassessed the validity of the terminal-velocity paradigm beyond the idealized assumptions of fixed cloud properties. 
To this end, we developed a generalized analytical framework for cloud motion under gravity, ram-pressure drag, mass exchange, and geometric evolution, and tested it against 3D hydrodynamical simulations including adiabatic evolution, radiative cooling, and thermal conduction.

Our main conclusions are as follows:
\begin{enumerate}
    \item[1.] In the constant-property limit, the terminal velocity remains a well-defined local equilibrium of the EOM, but it is generally \emph{not} a global attractor of the cloud dynamics.
    Whether it is approached depends on the ordering of the drag, free-fall, and convergence time-scales. Low-column-density clouds converge rapidly towards terminal motion, whereas dense clouds with large inertia per unit area can remain quasi-ballistic over much of their trajectories and approach the terminal regime only shortly before reaching the Galactic disc.

    \item[2.] The classical terminal-velocity picture is therefore valid only in a restricted dynamical regime, consistent with earlier qualitative expectations that cloud survival is a prerequisite for terminal-velocity applicability \citep{Benjamin:1999}. 
    It provides a useful description when the cloud remains sufficiently coherent and the convergence time is short compared to the time-scales of structural and thermal evolution. 
    It breaks down once mass exchange, deformation, and fragmentation alter the cloud properties on comparable or shorter time-scales than the global motion.

    \item[3.] We derived analytical solutions expressible in quadrature for constant-property clouds moving under quadratic drag in vertically varying Galactic density and gravitational profiles. 
    These solutions generalize the framework of \citetalias{Benjamin:1997} and provide a useful reference for interpreting cloud trajectories in a realistic stratified Milky Way background.

    \item[4.] We identified Bernoulli-driven lateral expansion as an additional dynamical effect not included in previous analytical terminal-velocity models. 
    This deformation increases the effective drag cross section and feeds back on the cloud motion already during the early coherent phase. 
    Our analytical expansion model reproduces the initial radius evolution in the simulations, but breaks down once the assumptions of quasi-steady flow and coherent cloud geometry are no longer satisfied.

    \item[5.] The hydrodynamical simulations show that adiabatic clouds are rapidly disrupted by the combined action of KH and RT instabilities, enhanced drag, and geometric deformation. 
    In this regime, the analytical and semi-analytical models provide only an early-time description, and the terminal-velocity paradigm loses predictive power once the cloud is disrupted and ceases to exist as a coherent object.

    \item[6.] Radiative cooling qualitatively changes the evolution.
    Condensation in the turbulent mixing layers leads to sustained mass growth, wake focusing, and additional momentum loading through accretion drag. 
    Under these conditions, the cloud remains dynamically coupled to the ambient medium and stays close to a terminal-velocity-like state for an extended period. 
    In this sense, cooling can \emph{restore} the practical relevance of the terminal-velocity concept even when the cloud no longer satisfies the assumptions of the constant-property model.

    \item[7.] Thermal conduction does not significantly alter the global dynamics in the present setup, but it modifies the internal structure of the wake. 
    In particular, it smooths temperature and density gradients, suppresses small-scale thermal fragmentation, and produces broader and less coherent condensations. Classical conductive evaporation remains dynamically negligible in our simulations.

    \item[8.] The semi-analytical model, based on our phenomenological mass-exchange formulation, captures the global evolution well whenever mass exchange is represented correctly. 
    Its predictive power is therefore fundamentally tied to the treatment of stripping, condensation, and evaporation, rather than to a detailed description of the instantaneous drag coefficient or cloud geometry. 
    This suggests that mass evolution is the dominant missing ingredient in simplified terminal-velocity models.

    \item[9.] Our simulations provide, to our knowledge, the first direct measurement of an effective drag coefficient for an infalling cloud in a stratified Galactic background medium. 
    The inferred coefficient is of order unity but strongly time-dependent, with typical values $\overline{C_\mathrm{d}}\sim 2$--3 depending on the adopted thermal physics. 
    This variability reflects the evolving cloud morphology and cautions against interpreting $C_\mathrm{d}$ as a fixed parameter in dynamical distance estimates. 
    It arises from substantial, time-dependent deformations of the cloud driven by hydrodynamical (Bernoulli-type) forces and ongoing mass exchange with the ambient medium, both of which lead to significant deviations from the commonly assumed value $C_\mathrm{d}=1$.

    \item[10.] The simulated cloud evolution leaves characteristic observational signatures. 
    In particular, the cooling runs produce VBs in position--velocity space, modest but structured extinction associated with mixed and condensed gas, and enhanced SXR emission tracing compression of the ambient medium. 
    These diagnostics provide a direct link between cloud dynamics and observable signatures of interaction at the disc--halo interface.
\end{enumerate}

Overall, our results show that the terminal-velocity paradigm remains most useful as a \emph{local and conditional} description of halo-cloud motion, rather than as a universal global law. 
It succeeds when clouds remain dynamically coherent or are kept strongly coupled to their environment by radiative condensation, but fails once deformation, mass loss, and fragmentation dominate the evolution. 
The limits of the terminal-velocity paradigm for IVCs and HVCs are therefore set not only by the Galactic background medium, but equally by the cloud's own structural and thermal evolution.
More generally, the framework developed here is not specific to IVCs and HVCs: once self-gravity and dark matter are included, and provided tidal disruption is negligible, it may equally describe other gravitational accretion systems -- such as infalling dwarf galaxies or compact gas clumps -- connecting our results to the broader context of galaxy evolution.

\section*{Acknowledgements}

We are grateful to the referee, Robert A.~Benjamin, for a careful and constructive report that substantially improved both the scope and the clarity of this paper.
We thank Consuelo L.~Guzman and Jan Bolte for their valuable preliminary work \citep{Guzman:2015,Guzman:2019}.
This research has made use of the \textsc{yt} astrophysics analysis software suite \citep{Turk:2011}, \textsc{matplotlib} \citep{Hunter:2007}, \textsc{numpy} \citep{Harris:2020}, and \textsc{scipy} \citep{Virtanen:2020} whose communities we thank for continued development and support.


\section*{Author contributions}

MMS conceived the project, developed the (semi-)analytical framework in its final form, performed the numerical simulations, carried out the data analysis, and wrote the manuscript.
DB contributed to the analytical development, including early extensions of \citetalias{Benjamin:1997} and the introduction of the Bernoulli-expansion concept.
JK advised on the observational diagnostics and contributed to drafting the corresponding sections.
All authors commented on the manuscript.

\section*{Data Availability}

The data underlying this article will be shared on reasonable request to the corresponding author.
The simulation movies are available as online supplementary material.



\bibliographystyle{mnras}
\bibliography{hvc} 



\appendix

\section{Analytical solution of the cloud EOM}
\label{app:solution}

In this appendix, we derive the analytical solution of the EOM, expressed in quadrature, in the dense-cloud limit ($\chi \gg 1$), for which buoyancy effects are negligible, under the assumption of constant cloud properties (mass, volume, frontal area, and drag coefficient).
In this case the specific mass-exchange rate $\beta = 0$, and the deceleration parameter $\alpha$ varies with height only through the background density.
Starting from the resulting EOM,
\begin{equation}
    \ode{\varv}{t} = \pm\alpha(z(t)) \varv^2(z(t)) - \varg(z(t)) ,
    \label{eq:eom}
\end{equation}
we seek a solution satisfying the initial condition $\varv(z(t_\mathrm{i})) = \varv(z_\mathrm{i}) \equiv \varv_\mathrm{i}$. To this end, we first introduce a new variable,
\begin{equation}
    \vary(z(t)) \equiv \left(\ode{z}{t}\right)^{\!2} = \varv^2(z(t)) ,
    \label{eq:newvar}
\end{equation}
which we differentiate with respect to $t$ to obtain
\begin{equation}
    \begin{split}
        \ode{\vary}{t} &= 2 \varv(z(t)) \ode{\varv}{t}\\
        &= 2 \varv(z(t)) \left[\pm\alpha(z(t)) \varv^2(z(t)) - \varg(z(t))\right] ,
    \end{split}
    \label{eq:dydt}
\end{equation}
where we have used equation~\eqref{eq:eom} in the last step. Since by the chain rule $\mathrm{d}\vary/\mathrm{d}t = (\mathrm{d}\vary / \mathrm{d}z)(\mathrm{d}z / \mathrm{d}t) = \varv \mathrm{d}\vary / \mathrm{d}z$, we can rewrite equation~\eqref{eq:dydt} as
\begin{equation}
    \ode{\vary}{z}=2 \left[\pm\alpha(z) \varv^2(z) - \varg(z)\right] ,
\end{equation}
or, equivalently,
\begin{equation}
    \ode{\vary}{z} = 2 \left[\pm\alpha(z) \vary(z) - \varg(z)\right] .
\end{equation}

We have thus transformed the original non-linear first-order ordinary differential equation (ODE) in time into a linear first-order ODE in $z$. 
Multiplying both sides by the integrating factor
\begin{equation}
    \mu(z)=\e^{\mp 2 \int_{z_\mathrm{i}}^z \alpha(z') \dif z'}
\end{equation}
yields
\begin{equation}
\begin{split}
    &\e^{\mp 2 \int_{z_\mathrm{i}}^z \alpha(z') \dif z'} \ode{\vary}{z} \mp 2 \alpha(z) \e^{\mp 2 \int_{z_\mathrm{i}}^z \alpha(z') \dif z'} \vary(z)\\
    &\quad = -2 \varg(z) \e^{\mp 2 \int_{z_\mathrm{i}}^z \alpha(z') \dif z'} .
\end{split}
\end{equation}
Applying the product rule to the left-hand side, this can be rewritten as
\begin{equation}
    \ode{}{z} \left[\vary(z) \e^{\mp 2 \int_{z_\mathrm{i}}^z \alpha(z') \dif z'}\right] = -2 \varg(z) \e^{\mp 2 \int_{z_\mathrm{i}}^z \alpha(z') \dif z'} .
\end{equation}
Integrating both sides with respect to height,
\begin{equation}
\begin{split}
    &\int_{z_\mathrm{i}}^z \ode{}{\tilde{z}} \left[\vary(\tilde{z}) 
    \e^{\mp 2 \int_{z_\mathrm{i}}^{\tilde{z}} \alpha(z') \dif z'}\right] \dif \tilde{z}\\
    &\quad = -2 \int_{z_\mathrm{i}}^z \varg(\tilde{z}) \e^{\mp 2 \int_{z_\mathrm{i}}^{\tilde{z}} \alpha(z') \dif z'} \dif \tilde{z} ,
\end{split}
\end{equation}
evaluates into
\begin{equation}
\begin{split}
    &\vary(z) \e^{\mp 2 \int_{z_\mathrm{i}}^z \alpha(z') \dif z'} - \vary(z_\mathrm{i})\\
    &\quad = -2 \int_{z_\mathrm{i}}^z \varg(\tilde{z}) \e^{\mp 2 \int_{z_\mathrm{i}}^{\tilde{z}} \alpha(z') \dif z'} \dif \tilde{z} .
\end{split}
\end{equation}
Solving for $\vary(z)$, we get
\begin{equation}
\begin{split}
    \vary(z) &= \e^{\pm 2 \int_{z_\mathrm{i}}^z \alpha(z') \dif z'}\\
    &\quad \times \left[\vary(z_\mathrm{i}) - 2 \int_{z_\mathrm{i}}^z \varg(\tilde{z})
    \e^{\mp 2 \int_{z_\mathrm{i}}^{\tilde{z}} \alpha(z') \dif z'} \dif \tilde{z}\right] .
\end{split}
\end{equation}
Substituting equation~\eqref{eq:newvar} with $\vary(z_\mathrm{i}) = \varv_\mathrm{i}^2$, and taking the square root of both sides, we arrive at the expression presented as equation~\eqref{eq:eomsols}.

\section{Derivation of the flow field around the cloud}
\label{app:flowfield}

To quantify the pressure differences underlying the Bernoulli effect discussed in Section~\ref{sec:bernoulli}, we derive the velocity field around a dense spherical cloud of radius $R_\mathrm{cl}$ moving with velocity $\varv$ through a dilute ambient medium at rest.
Under the assumptions of steady, incompressible, and irrotational flow, this problem admits an analytical solution.

In the cloud's rest frame, the ambient gas flows uniformly past the cloud with velocity $-\varv$ along the Galactic $z$-axis. 
The velocity field $\mathbfit{u}$ can therefore be written as the gradient of the velocity potential $\phi$, which satisfies the Laplace equation,
\begin{equation}
    \nabla^2\phi = 0 .
\end{equation}

In spherical coordinates $(r,\theta,\varphi)$, the general axisymmetric solution to the Laplace equation is
\begin{equation}
    \phi(r,\theta) = \sum_{n=0}^\infty \left(A_n r^n + \frac{B_n}{r^{n+1}}\right) P_n(\cos\theta) ,
\end{equation}
where $P_n(\cos\theta)$ are the Legendre polynomials, and $A_n$ and $B_n$ are constants.
For uniform flow past a cloud, only the dipole term ($n=1$) contributes: the monopole term ($n=0$) corresponds to a spherically symmetric source or sink, while terms with $n \ge 2$ represent higher multipoles absent in the present geometry.
Thus, the general solution simplifies to
\begin{equation}
    \phi(r,\theta) = \left(A_1 r + \frac{B_1}{r^2}\right) \cos\theta ,
\label{eq:laplacesol}
\end{equation}
upon noting that $P_1(\cos\theta) = \cos\theta$.

Boundary conditions determine $A_1$ and $B_1$.
The far-field condition requires the flow to approach uniformity along the $z$-direction as $r\rightarrow\infty$, implying
\begin{equation}
    \phi \rightarrow -\varv r \cos\theta
    \qquad \textrm{as } r\rightarrow\infty .
\end{equation}
Comparison with equation~\eqref{eq:laplacesol} yields
\begin{equation}
    A_1 = -\varv .
\end{equation}

Because the cloud is much denser than the surrounding medium, it may be approximated as a rigid, impermeable obstacle to the flow.
The no-penetration boundary condition therefore requires the radial velocity,
$u_r=\partial\phi/\partial r$, to vanish everywhere on the cloud surface, i.e.
\begin{equation}
    u_r(R_\mathrm{cl},\theta)=0
    \qquad \text{for all }\theta .
\end{equation}
Differentiating equation~\eqref{eq:laplacesol} with respect to $r$ and evaluating at the cloud surface gives
\begin{equation}
    u_r(R_\mathrm{cl},\theta)
    =
    \left(
    A_1-\frac{2B_1}{R_\mathrm{cl}^3}
    \right)\cos\theta .
\end{equation}
Since this must vanish for all $\theta$, the prefactor must satisfy
\begin{equation}
    A_1-\frac{2B_1}{R_\mathrm{cl}^3}=0 ,
\end{equation}
which implies
\begin{equation}
    B_1=-\frac{\varv R_\mathrm{cl}^3}{2} .
\end{equation}

The velocity potential therefore becomes
\begin{equation}
    \phi(r,\theta)
    =
    -\varv
    \left(
    r+\frac{R_\mathrm{cl}^3}{2r^2}
    \right)
    \cos\theta .
\end{equation}
The corresponding velocity components are
\begin{align}
    u_r
    &=
    \frac{\partial\phi}{\partial r}
    =
    -\varv
    \left(
    1-\frac{R_\mathrm{cl}^3}{r^3}
    \right)
    \cos\theta ,
    \\
    u_\theta
    &=
    \frac{1}{r}\frac{\partial\phi}{\partial\theta}
    =
    \varv
    \left(
    1+\frac{R_\mathrm{cl}^3}{2r^3}
    \right)
    \sin\theta .
\end{align}
The squared velocity magnitude,
$u^2=u_r^2+u_\theta^2$, is therefore
\begin{equation}
u^2
=
\varv^2
\left[
\left(
1-\frac{R_\mathrm{cl}^3}{r^3}
\right)^{\!2}
\cos^2\theta
+
\left(
1+\frac{R_\mathrm{cl}^3}{2r^3}
\right)^{\!2}
\sin^2\theta
\right] .
\label{eq:sqvelmag}
\end{equation}

\section{Resolution study}
\label{app:resostudy}
\begin{figure}
    \includegraphics[width=\columnwidth]{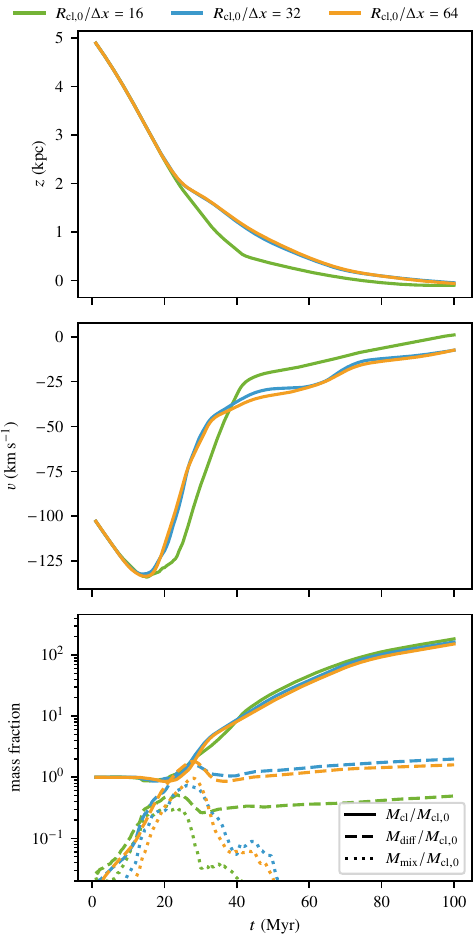}
    \caption{
    Time evolution of the cloud's COM vertical position (top panel), COM vertical velocity (middle panel), and the mass of the bulk, diffuse, and mixing components, each normalized to the initial cloud mass (bottom panel), for the RC+TC simulations at three resolutions corresponding to 16 (green), 32 (blue; fiducial), and 64 (orange) finest-grid cells per initial cloud radius.
    }
    \label{im:resostudy}
\end{figure}
We assess numerical convergence using the RC+TC setup, comparing runs with 16, 32, and 64 cells per initial cloud radius (Fig.~\ref{im:resostudy}).

The global dynamics are well converged. The COM trajectory and velocity evolution agree closely between the 32- and 64-cell runs, both showing a consistent monotonic infall. The 16-cell run follows the same overall behaviour, with deviations becoming apparent only at later stages of the evolution.

The mass evolution reveals a more selective dependence on resolution. The bulk-cloud mass is remarkably robust, with all runs exhibiting nearly identical growth, indicating that the net condensation onto the dense phase is largely insensitive to resolution. In contrast, the diffuse and mixing components show systematic differences: both are reduced in the 16-cell run, while the 32- and 64-cell runs are in much closer agreement. This indicates that the multiphase structure of the cloud and its surrounding interface is under-resolved at low resolution.

The late-time deviations in the COM evolution are consistent with these differences. While the total mass of the dense cloud is similar, variations in the amount and structure of diffuse and mixed gas imply differences in cloud morphology and effective cross section, which in turn affect the coupling to the background medium.

We therefore conclude that the fiducial resolution (32 cells per initial cloud radius) provides a reliable description of the global dynamics and bulk-cloud evolution. Lower resolution reproduces the overall behaviour but underestimates the diffuse and mixed gas and shows larger deviations at late times, so that quantitative conclusions on phase structure and late-time dynamics require at least the fiducial resolution.

\bsp	
\label{lastpage}
\end{document}